\documentclass[
  journal=pasa,
  manuscript=research-paper, 
  year=202X,
  volume=YY,
  hyperref,
]{cup-journal}

\usepackage{amsmath}
\usepackage{amssymb,microtype,siunitx,booktabs}
\usepackage{CJKutf8}
\newcommand{\chinese}[1]{\begin{CJK*}{UTF8}{gbsn}#1\end{CJK*}}

\usepackage{xcolor}
\newcommand{\updated}[1]{#1}

\newcommand{\updatednew}[1]{#1}

\newcommand{\lptone}{ASKAP~J165130.3$-$450520}
\newcommand{\lpttwo}{ASKAP~J170036.6$-$445758}
\newcommand{\unkone}{ASKAP~J045522.4$-$350523}

\newcommand{\unkthree}{ASKAP~J212758.7$-$470528}

\newcommand{\unkthreegx}{GLEAM-X~J212758.5$-$470524}

\hypersetup{
    colorlinks=true,
    citecolor=blue,  
    urlcolor=blue, 
    linkcolor=blue, 
    }

\title{VASTER: The ASKAP real-time fast-imaging pipeline -- overview and discovery of two long period transients}

\author{Yuanming Wang (\chinese{王远明})}
\affiliation{Centre for Astrophysics and Supercomputing, Swinburne University of Technology, Hawthorn, VIC 3122, Australia}
\alsoaffiliation{ARC Centre of Excellence for Gravitational Wave Discovery (OzGrav), Australia}
\email[Y. Wang]{yuanmingwang@swin.edu.au}

\author{Dougal Dobie}
\affiliation{Sydney Institute for Astronomy, School of Physics, The University of Sydney, NSW 2006, Australia}
\alsoaffiliation{ARC Centre of Excellence for Gravitational Wave Discovery (OzGrav), Australia}

\author{Tara~Murphy}
\affiliation{Sydney Institute for Astronomy, School of Physics, The University of Sydney, NSW 2006, Australia}
\alsoaffiliation{ARC Centre of Excellence for Gravitational Wave Discovery (OzGrav), Australia}

\author{Emil Lenc}
\affiliation{Australia Telescope National Facility, CSIRO Space and Astronomy, PO Box 76, Epping, NSW 1710, Australia}

\author{David L. Kaplan}
\affiliation{Department of Physics \& Astronomy, University of Wisconsin-Milwaukee, P.O. Box 413, Milwaukee, WI 53201, USA}
\alsoaffiliation{ARC Centre of Excellence for Gravitational Wave Discovery (OzGrav), Australia}

\author{Joshua Pritchard}
\affiliation{Australia Telescope National Facility, CSIRO Space and Astronomy, PO Box 76, Epping, NSW 1710, Australia}

\author{Daniel Mitchell}
\affiliation{Australia Telescope National Facility, CSIRO Space and Astronomy, PO Box 76, Epping, NSW 1710, Australia}

\author{Wasim Raja}
\affiliation{Australia Telescope National Facility, CSIRO Space and Astronomy, PO Box 76, Epping, NSW 1710, Australia}

\author{Matthew Whiting}
\affiliation{Australia Telescope National Facility, CSIRO Space and Astronomy, PO Box 76, Epping, NSW 1710, Australia}

\author{Owen Cole}
\affiliation{Astronomy Data and Computing Services (ADACS), The Centre for Astrophysics \& Supercomputing, Swinburne University of Technology, Australia}

\author{Paul J. Hancock}
\affiliation{Curtin Institute for Data Science, Curtin University, Bentley, WA 6102, Australia}
\alsoaffiliation{Astronomy Data and Computing Services, Curtin University, Bentley, WA 6102, Australia}

\author{Jiting Hu}
\affiliation{Astronomy Data and Computing Services (ADACS), The Centre for Astrophysics \& Supercomputing, Swinburne University of Technology, Australia}

\author{Yu Wing Joshua Lee}
\affiliation{Sydney Institute for Astronomy, School of Physics, The University of Sydney, NSW 2006, Australia}
\alsoaffiliation{Australia Telescope National Facility, CSIRO Space and Astronomy, PO Box 76, Epping, NSW 1710, Australia}
\alsoaffiliation{ARC Centre of Excellence for Gravitational Wave Discovery (OzGrav), Australia}

\author{Alex Massen-Hane}
\affiliation{Curtin Institute for Data Science, Curtin University, Bentley, WA 6102, Australia}
\alsoaffiliation{Astronomy Data and Computing Services, Curtin University, Bentley, WA 6102, Australia}

\author{Shibli Saleheen}
\affiliation{Astronomy Data and Computing Services (ADACS), The Centre for Astrophysics \& Supercomputing, Swinburne University of Technology, Australia}

\author{Raymond Shao}
\affiliation{Sydney Institute for Astronomy, School of Physics, The University of Sydney, NSW 2006, Australia}

\author{Lei Zhang}
\affiliation{State Key Laboratory of Radio Astronomy and Technology, National Astronomical Observatories, Chinese Academy of Sciences, Beijing 100101, China}
\alsoaffiliation{Centre for Astrophysics and Supercomputing, Swinburne University of Technology, Hawthorn, VIC 3122, Australia}

\author{Adarsh Bathula}
\affiliation{Centre for Astrophysics and Supercomputing, Swinburne University of Technology, Hawthorn, VIC 3122, Australia}

\author{Manisha Caleb}
\affiliation{Sydney Institute for Astronomy, School of Physics, The University of Sydney, NSW 2006, Australia}
\alsoaffiliation{ARC Centre of Excellence for Gravitational Wave Discovery (OzGrav), Australia}

\author{Raghav Girgaonkar}
\affiliation{Department of Physics \& Astronomy, University of Wisconsin-Milwaukee, P.O. Box 413, Milwaukee, WI 53201, USA}

\author{Ashna Gulati}
\affiliation{Sydney Institute for Astronomy, School of Physics, The University of Sydney, NSW 2006, Australia}
\alsoaffiliation{ARC Centre of Excellence for Gravitational Wave Discovery (OzGrav), Australia}
\alsoaffiliation{Australia Telescope National Facility, CSIRO Space and Astronomy, PO Box 76, Epping, NSW 1710, Australia}

\author{Natasha Hurley-Walker}
\affiliation{International Centre for Radio Astronomy Research, Curtin University, Bentley, WA, Australia}

\author{Iris de Ruiter}
\affiliation{Sydney Institute for Astronomy, School of Physics, The University of Sydney, NSW 2006, Australia}
\alsoaffiliation{ARC Centre of Excellence for Gravitational Wave Discovery (OzGrav), Australia}

\author{Ryan M. Shannon}
\affiliation{Centre for Astrophysics and Supercomputing, Swinburne University of Technology, Hawthorn, VIC 3122, Australia}
\alsoaffiliation{ARC Centre of Excellence for Gravitational Wave Discovery (OzGrav), Australia}

\author{Gregory R. Sivakoff}
\affiliation{Department of Physics, University of Alberta, CCIS 4-181, Edmonton, AB T6G 2E1, Canada}

\received {dd Mmm YYYY}
\revised  {dd Mmm YYYY}
\accepted {dd Mmm YYYY}
\published{22 September 202X}


\begin{document}

\begin{abstract}
Recent developments in widefield radio telescopes have enabled searches of a new region of parameter space in the time domain: timescales of seconds to minutes, that have been overlooked in traditional surveys. These searches have revealed a new population of sources: long period transients, which typically show periodic behaviour of minutes to hours. In addition they have detected phenomena ranging from extreme scintillation to stellar radio bursts. However, almost all searches to date have involved archival data that has been processed in offline, batch mode. In this context, we present VASTER, the first short-timescale imaging and transient detection pipeline running in real time on a widefield radio telescope. VASTER has been running on the Australian SKA Pathfinder (ASKAP) since July 2025, and images most of the ASKAP survey project data on timescales of 15 minutes. In this paper we describe the VASTER system, and present the results from the first two weeks of operation, including the discovery of two long period transients: \lptone{} with a period of 6.48 hours and \lpttwo{} with a period of 4.69 hours. The detection of these two sources adds to the small, but growing, population of long period transients, as well as demonstrating the potential of VASTER to explore this region of transient parameter space.
\end{abstract}

\section{Introduction}
\label{sec:introduction}
Radio sources show variability on timescales spanning orders of magnitude, from nanosecond bursts from the Crab pulsar \citep{Hankins2003Natur.422..141H} through to the slow evolution of radio supernova over months to years \citep{Weiler2002ARA&A..40..387W}. 
However, the timescales of seconds to hours have been relatively unexplored by untargeted surveys \citep{Murphy2026PASA...43....6M}. 
This variability is too slow to be detectable with traditional single-dish techniques, but too fast to be detectable \updated{with the typical search cadences of traditional imaging surveys.}

In the past decade there have been a small number of projects that have conducted imaging transient searches on timescales of order \updated{10\,s}. 
These early attempts were mainly at low frequencies (tens of MHz) with \updated{low angular resolution and poor sensitivity} $\sim$tens of Jy \citep[e.g.,][]{Lazio2010AJ....140.1995L,Obenberger2015JAI.....450004O}. 


The advent of high-sensitivity, well-sampled (in the $u$-$v$ plane), widefield radio surveys has opened up this domain for exploration, and early searches have revealed a range of phenomena. 
Perhaps most notable are long period transients \updated{(LPTs; e.g., \citealt{Hurley-Walker2022Natur.601..526H,Hurley-Walker2023Natur.619..487H,Caleb2024NatAs...8.1159C})}
and slow pulsars \citep[e.g.,][]{Caleb2022NatAs...6..828C,Wang2025ApJ...982L..53W}, two emerging, and related classes. 
\updated{LPTs} are sources that show short, powerful, bursts of radio emission, that is \updated{periodic and often highly polarised}. 
There have been $\sim$14 detected to date\footnote{\url{https://vast-survey.org/LPTs/}}, exhibiting a wide range of observed properties, with periods ranging from minutes to hours, and \updated{long-term on-and-off activity windows} \citep[e.g.,][]{Hurley-Walker2022Natur.601..526H,Hurley-Walker2023Natur.619..487H,Hurley-Walker2024ApJ...976L..21H,Caleb2024NatAs...8.1159C,Dong2025ApJ...990L..49D,Wang2025Natur.642..583W,Lee2025NatAs...9..393L}. 
In addition to these new classes of sources, searches of these timescales also reveal rapidly scintillating active galactic nuclei, some of which have been observed in geometric alignment \citep{Wang2021MNRAS.502.3294W}, eclipsing pulsars \citep{Zic2024MNRAS.528.5730Z,Petrou2026PASA...43....7P}, and many types of radio stars \citep[e.g.][]{Driessen2024PASA...41...84D}. 

Motivated by these results, and the prospect of serendipitous discoveries, there have been a number of searches for short-timescale variability. 
While low-frequency instruments continue to play an important role in exploring this regime \citep[e.g.,][]{deRuiter2024MNRAS.531.4805D,Horvath2025PASA...42..129H}, short-timescale searches have increasingly been extended to mid-frequency ($\sim$GHz) facilities such as ASKAP and MeerKAT. 
On MeerKAT, \citet{Fijma2024MNRAS.528.6985F} searched for variability on 8\,s, 128\,s, and 1\,h timescales in the field of NGC~5068, but found no transient candidates.
On ASKAP, \cite{Wang2021MNRAS.502.3294W} and \cite{Wang2023MNRAS.523.5661W} performed transient \updated{searches} on a timescale of \updated{15\,min}, and detected a number of extreme scintillating galaxies, pulsars, and stellar flares. 
More recently, \cite{Lee2026MNRAS.545f2008L} performed a search on a timescale of \updated{10\,s} in ASKAP Galactic-plane fields targeting LPTs; while no new LPTs were discovered, several flaring stars were detected.
For all of these projects, the short timescale \updated{imaging} and transient detection was run in offline batch mode. This limits the opportunity for rapid follow-up of the transients detected, which is particularly important for 
\updated{source classes that can become quiescent rapidly.} 

ASKAP \citep{Hotan2021PASA...38....9H} can search for millisecond transients, such as fast radio bursts, in real time using the Commensal Real-time ASKAP Fast Transients (CRAFT) backend, which operates on time-domain data at $\sim$1\,ms resolution, incoherently adding intensities from individual antennas \citep{Macquart2010PASA...27..272M,Shannon2025PASA...42...36S}. 
The recent CRAFT coherent upgrade (CRACO) extends this capability and operates on timescales between $\sim$110\,ms and 3\,ms, 
\updated{using the correlated visibility data to generate a series of de-dispersed images of the mean-subtracted residual sky \citep{Wang2025PASA...42....5W}.}
While these systems have been highly successful in rapidly detecting millisecond-timescale transients \citep[e.g.,][]{Shannon2025PASA...42...36S}, a real-time fast-imaging system operating on standard image-domain data is crucial for studying slower variability (seconds to hours) in a timely manner.

\updated{A fast imaging pipeline has recently been developed for MeerKAT.} 
TRON (Transient Radio Observations for Newbies) images MeerKAT data \updated{at the raw correlator dump rate (e.g., 2\,s, 4\,s, or 8\,s, depending on the observing mode), and searches for radio transients on these or longer timescales using a tunable parameter \citep{Smirnov2025MNRAS.538L..62S}.} 
Searches of archival MeerKAT data with TRON have resulted in the detection of millisecond pulsars \citep{Smirnov2025MNRAS.538L..62S} and a stellar flare \citep{Smirnov2025MNRAS.538L..89S}. 
At the time of writing, work is underway to get TRON running routinely on MeerKAT observations (I. Heywood, {\it private communication}).

\updated{
The ASKAP Survey for Variables and Slow Transients (VAST; \citealt{Murphy2013PASA...30....6M}) is one of the key ASKAP Survey Science Projects\footnote{\url{https://www.atnf.csiro.au/projects/science/wide-area-surveys/askap-survey-science-projects/}}. 
VAST began its pilot survey in 2019 \citep{Murphy2021PASA...38...54M} and its full survey in 2022 \citep{deRuiter2026PASA...43...39D}.
The VAST transient pipeline \citep{Pintaldi2022ASPC..532..333P} is designed to identify radio variables and transients from the VAST surveys by associating sources across multiple epochs, with typical observing cadences ranging from days to months \citep{deRuiter2026PASA...43...39D}.
The VAST team is able to perform time-domain analysis commensally on other ASKAP Survey Science Projects. 
In contrast to the short, multi-epoch VAST observations, these surveys are typically longer ($\gtrsim$ a few hours) with limited or no repeat epochs, making them ideal for exploring faster, intra-observation transients. 
}

In this paper, we describe the `VASTER'\footnote{The name VASTER is a play on words: it is a `faster' version of the VAST transient detection pipeline \citep{Murphy2013PASA...30....6M,Murphy2021PASA...38...54M,Pintaldi2022ASPC..532..333P}.} 
real-time fast imaging pipeline. 
\updated{VASTER is primarily developed for long observations to identify intra-observation transients on timescales from seconds to hours. 
\citet{Wang2021MNRAS.502.3294W} demonstrated the prototype and concept of VASTER, while \citet{Wang2023MNRAS.523.5661W} described the workflow and technical details, and demonstrated its offline performance with ASKAP pilot surveys. 
As an extension of the previous work, VASTER was officially deployed on ASKAP in July 2025 and has been running as part of the standard ASKAP data processing workflow since then, imaging long observations on timescales of 15\,min. }
Transient candidates are made available to users even before the standard imaging data products are made available, making it the first such system to be routinely doing image-plane transient detection in \updated{real time}. 


In Section~\ref{sec:vaster} we describe the VASTER system and transient detection algorithm. In Section~\ref{sec:survey} we describe how VASTER operates commensally on the ASKAP surveys. In Section~\ref{sec:results} we present the results from the first two weeks of VASTER operations, including the discovery of two LPTs, along with a host of other transients. Finally, in Section~\ref{sec:discussion} we discuss our results and summarise our future plans for VASTER.

\section{VASTER system description}
\label{sec:vaster}

\begin{figure*}[t!]
\centering
\includegraphics[width=\linewidth]{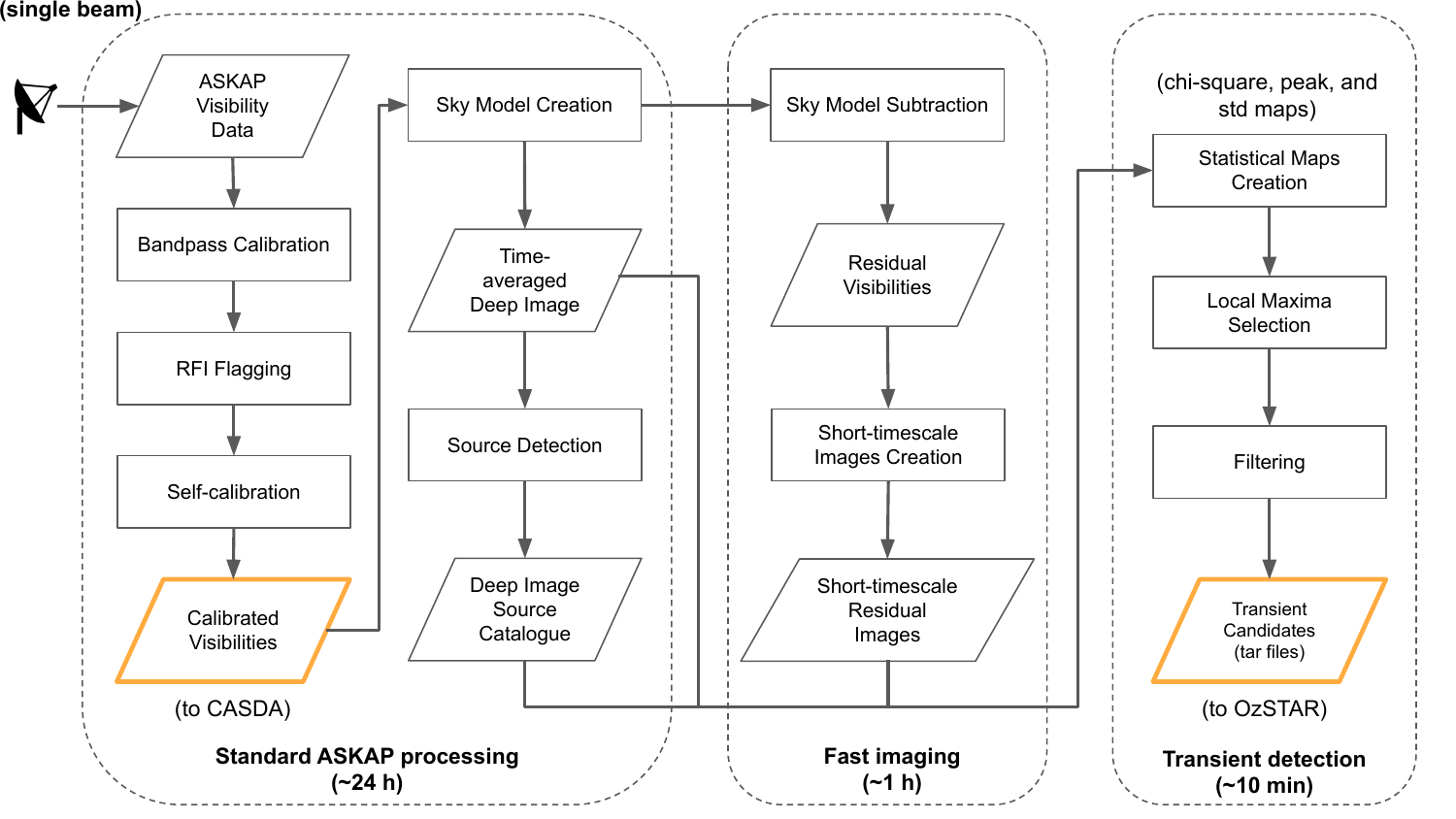}
\caption{Flowchart of the VASTER real-time system. 
An overview of the system is described in Section~\ref{sec:vaster}, with the \updated{`fast imaging'} described in Section~\ref{sec:fast_imaging_processing} and the `transient detection' described in Section~\ref{sec:transient_detection}.
All processing is performed independently on individual beams. 
\updated{We list the typical processing time for each part for a typical 10\,h observation.}
The inputs for the transient detection are the time-averaged deep image, the deep image source catalogue, and the short-timescale residual images. 
Two data products are saved during the process (marked as orange): the standard ASKAP calibrated visibilities (archived in CASDA) and the VASTER transient candidates (tar files; stored on OzSTAR). 
See more details about data storage and release in Section~\ref{sec:data_storage_and_release}. }
\label{fig:flowchart}
\end{figure*}

ASKAP consists of 36 fully steerable 12\,m dishes operating at radio frequencies of 700--1800\,MHz with instantaneous correlated bandwidths of 288\,MHz and maximum baselines of 6\,km \updated{\citep{Johnston2007PASA...24..174J,Hotan2021PASA...38....9H}}. 
The wide bandwidths, together with the number of antennas and their layout provides excellent instantaneous $(u,v)$ coverage, enabling high-quality snapshot imaging. 
Each antenna is equipped with a Phased Array Feed at its focal plane, elements of which are electronically combined to form 36 dual-polarised beams, providing a large instantaneous field of view. 
These specifications make ASKAP well-suited to discovering and localising transient and variable sources on sub-hour timescales. 

This short-timescale `fast imaging' capability has been demonstrated in previous work \citep[e.g.,][]{Wang2021MNRAS.502.3294W,Rigney2022MNRAS.516..540R,Wang2023MNRAS.523.5661W,Dobie2023MNRAS.519.4684D,Lee2026MNRAS.545f2008L}, but has been limited to dedicated offline processing of specific datasets available on the ASKAP archive, leading to substantial latency in transient discovery \updated{and preventing timely follow-up}. 

The VASTER real-time system presented here extends the existing ASKAPsoft workflow \citep{Hotan2021PASA...38....9H} to carry out sky model subtraction, intra-observation imaging, and a custom variability search. 
Figure \ref{fig:flowchart} describes how the system is integrated directly into the standard science processing and produces transient and variable candidates in \updated{real time} for human vetting. 

The standard ASKAPsoft data processing workflow is described by \citet{Hotan2021PASA...38....9H}. 
In the rest of this section, we describe the key components of the VASTER system. 
\updated{
Note that the VASTER real-time system presented here largely follows the offline processing workflow described by \citet{Wang2023MNRAS.523.5661W}. 
The key difference is that the real-time implementation makes use of existing functionality available in the ASKAPsoft workflow to improve efficiency (see Section~\ref{sec:fast_imaging_processing}). 
}

\subsection{\updated{Fast imaging}}
\label{sec:fast_imaging_processing}
For all observations, visibility data from the ASKAP hardware correlator are written in parallel to the Pawsey\footnote{\url{https://pawsey.org.au/}} supercomputing centre, \updated{with each PAF beam stored in its own CASA measurement set file; that is, a total of 36 measurement sets for one observation.
} 
The high resolution spectral line datasets are further partitioned into 6 frequency chunks \citep{Voronkov2020EPJWC.24501038V}. 
The ASKAPsoft pipeline reads these visibilities and then carries out bandpass calibration, flagging, spectral averaging to 1\,MHz resolution (for high spectral resolution data), self-calibration and imaging of these continuum averaged data for the entire observation as described in \citet{Hotan2021PASA...38....9H}. It produces calibrated per-beam visibilities along with the resulting mosaiced `deep' images and associated source-finder output that are uploaded to the CSIRO ASKAP Data Archive (CASDA). However, the per-beam sky models are not archived, meaning that the offline versions of the ASKAP fast imaging process used in the literature \citep[e.g.,][]{Wang2023MNRAS.523.5661W} have had to download the data from CASDA and re-image the calibrated visibilities in order to obtain sky models for subtraction. 

This duplication of work is computationally inefficient, but more importantly introduces substantial latency in finding transients. The aim of the VASTER project was to incorporate fast imaging as part of the standard science processing workflow to remove this inefficiency. 
\updated{It also removes other sources of delay, such as the time required for data to become publicly available in the archive, which can take weeks to months due to scientific validation by the survey science teams.}

As part of the VASTER extension to the ASKAP pipelines, \updated{all processing is performed \updatednew{on the Setonix}\footnote{\url{https://pawsey.org.au/systems/setonix/}} supercomputer housed at Pawsey, with each beam processed independently throughout.} 
We first create per-beam residual visibilities by subtracting the existing sky model from the calibrated visibilities using \updated{the ASKAPsoft application \textit{ccontsubtract}\footnote{\url{https://www.atnf.csiro.au/computing/software/askapsoft/sdp/docs/current/calim/ccontsubtract.html}}.} 
We then produce intra-observation residual images \updated{by splitting and imaging the residual visibilities at 15\,min intervals using the ASKAPsoft \textit{imager}\footnote{\url{https://www.atnf.csiro.au/computing/software/askapsoft/sdp/docs/current/calim/imager.html}} application.} 
The imaging parameters for these residual visibilities are treated separately to the deep imaging parameters. 
Since most of the continuum emission has been subtracted out, \updatednew{only dirty images are produced with no further deconvolution}.
The imaging uses multi-frequency synthesis but with no Taylor-term imaging. 
The cell-size parameter (oversampling of the synthesised beam) is kept at 2\,arcsec. The image shape for each beam is set to $\sim 2^{\circ}\times 2^{\circ}$ and matches the field-of-view of a single beam at the lowest ASKAP frequencies. 

\updated{The choice of timescale was dictated by several practical factors, including the I/O limitations on Setonix, as well as the computational and storage overheads associated with imaging on shorter time intervals and restrictions on the maximum number of files and concurrent jobs. As VASTER operates commensally within the standard ASKAP processing workflow, it is important to ensure that it does not adversely impact other survey pipelines running on the same system, and the current configuration therefore represents a conservative, pilot implementation.}
For example, using the native 10\,s resolution would produce 3600 images per beam for a standard 10\,h observation, and the use of separate imaging jobs for each image would lead to a high degree of overheads \updated{due to repeated fixed tasks (e.g., allocation of grids and definition of convolution functions), resulting in the transient imaging taking longer than the rest of the processing.}
While a variety of transient classes occur on much shorter timescales and hence require higher time resolution data to completely understand, as we demonstrate in this work, some are still detectable with coarse light curves. 
\updated{Future improvements to the processing workflow (e.g., a properly parallelised transient-imager) would remove these overheads, allowing higher time resolution imaging down to 10\,s.}
For a typical 10\,h observation, the current workflow produces about 40 15\,min residual images \updated{per beam (each image is about 50\,MB)}, with a typical rms noise of $\sim$200\,$\mu$Jy\,beam$^{-1}$ per image.



\subsection{Transient detection} 
\label{sec:transient_detection}

\updated{The fast imaging described above makes use entirely of existing ASKAPsoft functionality.
As ASKAPsoft lacks transient detection capability, we have developed a dedicated transient detection algorithm, which is containerised and installed as a module on Setonix. 
The underlying logic of the algorithm follows that used in the offline processing, as described in detail by \citet{Wang2023MNRAS.523.5661W}. 
Here, we summarise the main steps, including the selection criteria we use in the real-time system.}

The inputs \updated{to} the transient detection algorithm are the time-averaged deep image, the deep image source catalogue, and the short-timescale residual images.
The VASTER system generates three statistical maps from the \updated{short-timescale residual images} and \updated{identifies} transient candidates in the image plane. 

First, a data cube is constructed for each beam by stacking the series of \updated{short-timescale residual images}.
For a 10\,h observation, the cube size would be [40, 3584, 3584] pixels, where the first axis represents time and the other two represent spatial dimensions, \updated{i.e., in [time, dec, ra]}.
Three metrics\footnote{\updated{Note that \citet{Wang2023MNRAS.523.5661W} also created a `Gaussian map'.
However, subsequent offline analysis showed that this metric did not yield any additional genuine candidates and instead introduced a large number of false detections, and it was therefore not used in further processing.
This functionality remains available in the VASTER real-time system but is currently disabled, and may be reintroduced if it proves useful in future (e.g., for specific classes of transients or at different timescales).}} are calculated for each spatial pixel \updated{along the time axis}:
\begin{itemize}
    \item reduced chi-square ($\eta$); to produce a `chi-square map';
    \item peak signal-to-noise ratio ($\mathrm{S/N}_{\mathrm{peak}}$); to produce a `peak map';
    \item standard deviation ($\sigma_\mathrm{s}$); to produce a `standard-deviation map'.
\end{itemize}

The weighted reduced chi-square $\eta$ measures the significance of random variability and is defined as
\begin{equation}
\eta = \frac{1}{N - 1} \sum_{i=1}^{N} \frac{(S_i - \bar{S})^2}{\sigma_i^2},
\end{equation}
where $N$ is the total number of \updated{short-timescale residual images}, $S_i$ and $\sigma_i$ are the flux density and local rms noise of the $i$th residual image, and $\bar{S}$ is the weighted mean flux density.
\updated{The local rms noise is calculated from a square box with a size of 299 pixels around each spatial pixel in each short-timescale residual image.}
For $\mathrm{S/N}_{\mathrm{peak}}$, we calculate $S_i/\sigma_i$ for each spatial pixel and record the maximum value along the time axis.
We also calculate the weighted modulation index, defined as $m = \sigma_\mathrm{s}/\bar{S}$, to measure the magnitude of variability.

The VASTER system selects candidates as local maxima in either the chi-square or peak map, \updated{using a threshold of five times the standard deviation defined from the map distribution in logarithmic space to suppress extreme outliers.}
\updated{A minimum separation of 30 pixels is set between local maxima, such that peaks closer than this are merged into a single detection, reducing false detections from nearby bright sources.}
Candidates from both maps are combined and filtered using the following criteria:
\begin{enumerate}
    \item \updated{Sources located within 1.2 times the primary beam size (full width at half maximum) are selected to avoid regions near the beam edge, where the sensitivity decreases and the noise increases;}
    \item \updated{Sources with no counterpart in the deep image source catalogue within 20\,arcsec are selected as likely transients;}
    \item For sources with a close counterpart (separation $<$ 2\,arcsec), \updated{we selected only those with $m > 0.1$ (highly variable) and an integrated-to-peak flux ratio $<$ 1.5 (compact);}
    \item Sources with separations between 2 and 20\,arcsec are excluded as likely sidelobes or artefacts from the nearby source, unless the nearby source is faint ($<$ 2\,mJy).
\end{enumerate}

Once the processing is complete and the candidate list is finalised, VASTER automatically generates a light curve, deep-image cutout, and an animation of the short-timescale images for each candidate to help the team assess whether the candidate is real.
These data products are compressed into 36 tar files (one per beam). 
All objects in the final list are then passed to a web application for manual inspection.

\subsection{Candidate inspection tool}

\updated{The final outputs from VASTER are stored in a dedicated directory on Setonix and are automatically synchronised to a processing area on the OzSTAR\footnote{\url{https://supercomputing.swin.edu.au/}} supercomputer.}
We run a program on OzSTAR to routinely check the completeness of each observation (verifying that all 36 beams and final products are present).
For observations that pass this check, the program restructures the outputs and uploads them to the VASTER web application \updated{(see more details in~\ref{sec:vaster_outputs})}.
The system automatically assigns the inspection task for that SBID to an available team member, sends a notification to the team, and reports the number of candidates detected.

The web inspection tool is hosted on a virtual machine within the ARDC Nectar\footnote{\url{https://ardc.edu.au/services/ardc-nectar-research-cloud/}} Research Cloud and accessible to authorised users (currently, members of the VAST collaboration).
Users can filter assigned observation SBIDs and review each candidate through an interactive interface.
For each candidate, the web app displays key statistics (e.g., coordinates, $\eta$, $\mathrm{S/N}_{\mathrm{peak}}$, and $m$), the intra-observation light curve, a deep-image cutout, and an animation of \updated{short-timescale residual images}.
It also lists nearby sources from the local database (for candidates detected multiple times) and cross-matches with external \updatednew{databases} such as SIMBAD \citep{Wenger2000A&AS..143....9W} and the ATNF Pulsar Catalogue \citep{Manchester2005AJ....129.1993M}.
The tool enables team members to manually inspect each candidate to identify its source type (where possible) and rule out artefacts.
The typical number of transient candidates identified per observation is $\sim$5--20, of which $\sim$0--2 are confirmed as real astronomical objects after manual inspection.

\subsection{Latency}

\updated{Figure~\ref{fig:flowchart} lists the typical processing time for each part for a 10\,h ASKAP observation.}
The processing speed of the VASTER real-time system primarily depends on the sky-model creation step \updated{within the standard ASKAP processing}, which is the most time-consuming stage \updated{taking $\sim$15--20\,h}.
\updated{The other two parts (i.e., the VASTER extension) are typically much faster, taking $\sim$1\,h.} 
As VASTER is fully integrated into the main ASKAP workflow \updated{on Setonix}, no additional time is needed for data transfer or reprocessing.
\updated{However, Setonix is a shared supercomputing facility running all major ASKAP Survey Science Projects, and it is important to ensure that the VASTER system does not impact other surveys and that resources are balanced across projects.}
Continuum observations (e.g., EMU) are typically processed very promptly, while high spectral resolution observations (e.g., WALLABY and DINGO) can take longer if there are disk space shortages. 
\updated{As a result,}
the typical latency of the VASTER system is less than a few days, and at most $\sim$10 days, depending on the HPC load and disk space. 
This allows transient candidates to be identified well before the standard data products are validated and released, which can take weeks to months.

\subsection{Data storage and release}
\label{sec:data_storage_and_release}

The standard ASKAP science data products, including calibrated visibilities, mosaiced images (from all 36 beams), and catalogues, are deposited in CASDA.
These data are made available to public after quality control and validation by the survey science teams. 
VASTER transient \updated{candidates'} products, including their light curves, deep-image cutouts, and animations of the short-timescale images, are stored on OzSTAR and accessible to the VAST team members.
These VASTER products are not archived in CASDA.  
Other intermediate VASTER data products, including the time-averaged deep image (per beam), deep image source catalogue (per beam), \updatednew{and short-timescale residual images} (per beam), are not saved or archived to reduce storage usage. 
Future developments may include archiving model-subtracted visibilities or per-beam model images in CASDA, which would significantly reduce the offline reprocessing work when investigating variability on different timescales.

\section{Observations and survey strategy}
\label{sec:survey}

\begin{figure*}[t!]
\centering
\includegraphics[width=0.95\linewidth]{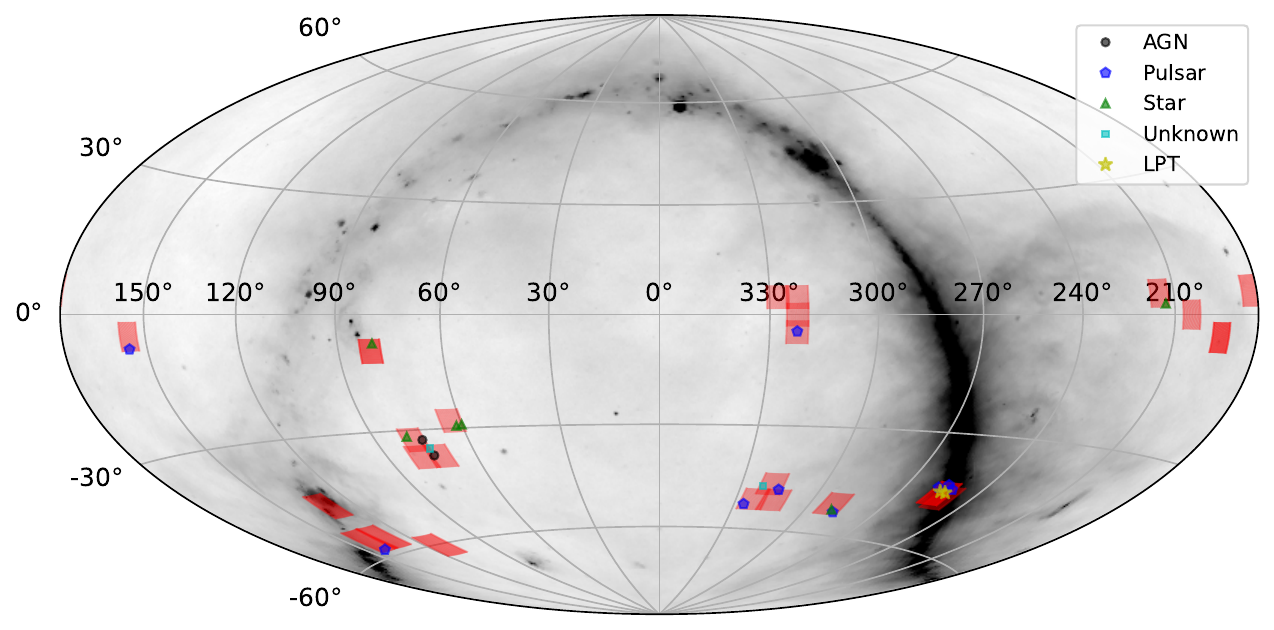}
\caption{Sky coverage of the observations used for the initial results (shown as red shaded regions). Markers represent detected transient and variable sources (see details in Section~\ref{sec:results}). The background is the diffuse Galactic radio emission at 943.5\,MHz modelled by \citet{Zheng2017MNRAS.464.3486Z}. }
\label{fig:survey_map}
\end{figure*}

\begin{table}[t!]
\centering
\caption{Details of the observations used in this work. Start times are in UTC, and $T$ represents the observation duration in hours. These observations are conducted at a central frequency of 943.5\,MHz with a bandwidth of 288\,MHz. }

\begin{tabular}{ccccr}
\toprule
SBID & RA & Dec & Start Time (UTC) & $T$ (h) \\
\midrule
74690 & 16:45:00.00 & $-$46:30:10.66 & 2025-07-11 08:58:00 & 10.0 \\
74712 & 09:18:00.00 & $-$55:43:29.41 & 2025-07-12 01:51:14 & 10.0 \\
74757 & 07:32:34.29 & $-$60:19:18.17 & 2025-07-13 00:05:02 & 10.0 \\
74787 & 04:48:00.00 & $-$37:14:54.56 & 2025-07-13 20:58:10 & 10.0 \\
74812 & 21:29:33.13 & $+$04:40:25.59 & 2025-07-14 15:05:57 & 5.0 \\
74815 & 05:22:23.28 & $-$09:20:42.19 & 2025-07-14 22:42:48 & 3.9 \\
74821 & 12:53:43.88 & $-$04:40:34.84 & 2025-07-15 08:04:50 & 5.1 \\
74824 & 21:00:00.00 & $-$51:07:06.40 & 2025-07-15 13:27:34 & 10.1 \\
74825 & 08:42:00.00 & $-$55:43:29.41 & 2025-07-15 23:31:51 & 10.1 \\
74856 & 12:53:43.88 & $-$04:40:34.84 & 2025-07-17 06:14:17 & 5.0 \\
74872 & 05:14:10.91 & $-$37:14:54.56 & 2025-07-17 20:05:43 & 10.0 \\
74873 & 10:23:17.01 & $-$04:40:34.84 & 2025-07-18 06:30:46 & 5.1 \\
74876 & 21:32:43.64 & $-$51:07:06.40 & 2025-07-18 14:02:19 & 10.0 \\
74878 & 08:45:00.00 & $-$46:30:10.66 & 2025-07-19 00:16:29 & 10.0 \\
74880 & 21:15:00.00 & $-$46:30:10.66 & 2025-07-19 12:32:02 & 10.0 \\
74881 & 05:22:23.28 & $-$09:20:42.19 & 2025-07-19 22:34:02 & 5.0 \\
74883* & 16:58:00.00 & $-$45:30:10.66 & 2025-07-20 07:42:38 & 4.3 \\
74945 & 14:19:42.09 & $+$04:40:25.59 & 2025-07-21 10:17:55 & 5.0 \\
74955* & 16:58:00.00 & $-$45:30:10.66 & 2025-07-22 07:19:12 & 10.1 \\
74957 & 04:12:00.00 & $-$27:58:10.20 & 2025-07-22 19:38:55 & 10.1 \\
74962 & 13:36:42.99 & $+$00:00:00.00 & 2025-07-23 06:38:12 & 5.1 \\
74967 & 21:29:33.13 & $+$00:00:00.00 & 2025-07-23 16:37:44 & 5.1 \\
75010 & 19:21:49.09 & $-$51:07:06.40 & 2025-07-24 11:16:59 & 10.0 \\
75059 & 21:29:33.13 & $-$04:40:34.84 & 2025-07-25 15:20:32 & 5.0 \\
75060 & 05:10:20.69 & $-$32:36:41.84 & 2025-07-25 20:23:51 & 10.1 \\
75061 & 12:10:44.78 & $+$04:40:25.59 & 2025-07-26 06:27:15 & 5.1 \\
75073 & 21:51:02.69 & $+$04:40:25.59 & 2025-07-26 14:52:15 & 5.0 \\
\bottomrule
\end{tabular}

\label{tab:survey}
\begin{tablenotes}[hang]
\item[*]ToO observations as follow ups of LPTs. 
\end{tablenotes}
\end{table}

ASKAP was designed primarily as a survey telescope, and has nine major survey science projects that take up $75\%$ of the time allocation. 
The VASTER system \updated{has been operating} commensally with these ASKAP science survey projects in real time since \updated{11~July~2025} (with the exception of CRAFT filler observations, which \updated{do} not use the standard processing pipeline). 
The current implementation produces images on \updated{15\,min} timescales, and therefore runs only on those surveys with long observations (greater than several hours).
These surveys are: 
\begin{enumerate}
    \item the Evolutionary Map of the Universe (EMU) survey \citep[][]{Norris2011PASA...28..215N,Norris2021PASA...38...46N}, a wide-area continuum survey aiming to map the entire southern sky at 943.5\,MHz with 10\,h \updated{of observing time} per field;
    
    \item the Widefield ASKAP L-band Legacy All-sky Blind surveY (WALLABY; \citealt{Koribalski2020Ap&SS.365..118K}), a survey of neutral hydrogen (H\,\textsc{i}) in the Local Universe aiming to observe half of the southern sky at 1367.5\,MHz with 8\,h \updated{of observing time} per field; and
    
    \item the Deep Investigation of Neutral Gas Origins (DINGO; \citealt{Rhee2023MNRAS.518.4646R}), which targets a small deep field with total \updated{observing} times of thousands of hours at 1367.5\,MHz.
\end{enumerate}
VASTER does not run on the Galactic ASKAP (GASKAP) survey \citep{Dickey2013PASA...30....3D}, as (i) the observations involve interleaving and hence do not produce a continuum data product in the same form as the other surveys; and (ii) the use of spectral zoom modes means that only a small fraction of the full bandwidth is available.
VASTER was initially operated on the First Large Absorption Survey in H\,\textsc{i} (FLASH; \citealt{Yoon2025PASA...42...88Y}), which aims to cover all sky with declination $<$+15$^\circ$ (excluding the Galactic plane) with 2\,h \updatednew{observations} at a central frequency of 855.5\,MHz to search for H\,\textsc{i} absorption. 
\updatednew{However, due to the high I/O load associated with per-beam source finding in the larger ($\sim$240\,GB), high-spectral-resolution FLASH cubes, the performance of VASTER is significantly reduced and can impact other survey processing running on Setonix.}
\updated{As a result,} VASTER has not been running on FLASH observations since \updated{28~August~2025} and will remain disabled until the ASKAP team is satisfied with improved performance.

In future versions of VASTER we aim to implement \updated{shorter} imaging timescales, down to 10\,s (the ASKAP sampling time).
\updated{This will allow VASTER to detect seconds-timescale transients (if sufficiently bright)} and to operate on short observations from surveys such as the Rapid ASKAP Continuum Survey \citep[RACS;][]{McConnell2020PASA...37...48M} and \updated{VAST} \citep{Murphy2013PASA...30....6M,Murphy2021PASA...38...54M}.

\section{Initial results}
\label{sec:results}

\begin{figure*}[t!]
\centering
\includegraphics[width=\linewidth]{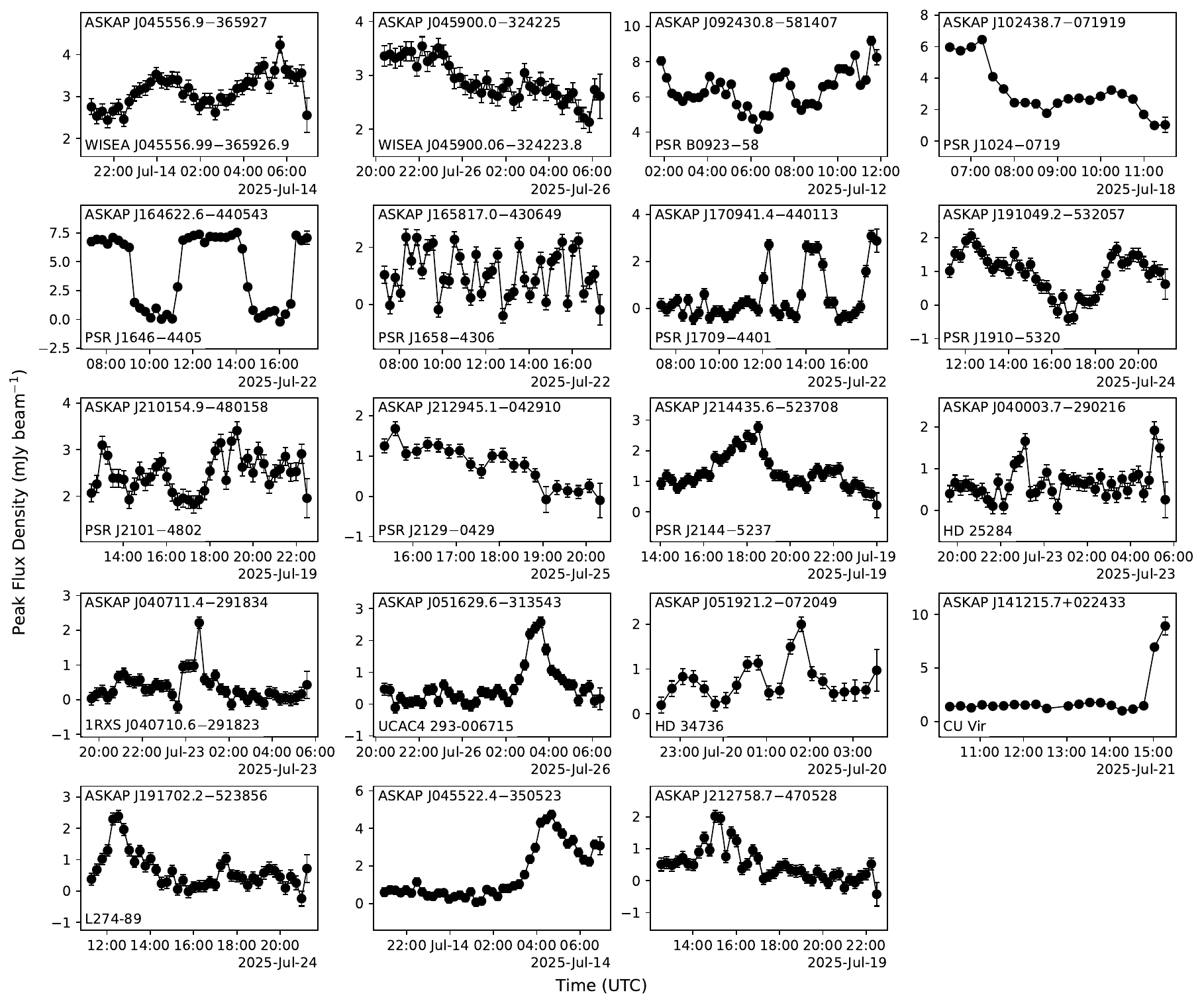}
\caption{Light curves of all transients (excluding the two LPTs, which are shown separately) in their detected epoch. Source identifications are given in the bottom-left corner of each panel. The two unidentified sources are shown on the bottom row, and have no label.}
\label{fig:lightcurves}
\end{figure*}

\begin{table*}[t!]
\centering
\caption{
Properties of detected transients. 
The table lists source coordinates (\updated{RA, Dec, and Galactic coordinates $l$ and $b$, in degrees}), variability metrics ($\eta$, $\mathrm{S/N}{\mathrm{peak}}$, and $m$), and the deep-image flux density ($S_\mathrm{deep}$; in mJy). 
For sources with multiwavelength counterparts, the corresponding SIMBAD or WISE identifications are provided. }

\begin{tabular}{lrrrrrrrrl}
\toprule
Name & RA & Dec & $l$ & $b$ & $\eta$ & $\mathrm{S/N}_{\mathrm{peak}}$ & $m$ & $S_\mathrm{deep}$ & Identification  \\
\midrule
AGN \\ \hline
ASKAP~J045556.9$-$365927 & 73.9871 & $-$36.9908 & 240.1425 & $-$38.1415 & 4.7 & 5.3 & 0.12 & 3.22 & WISEA J045556.99$-$365926.9 \\
ASKAP~J045900.0$-$324225 & 74.7499 & $-$32.7069 & 234.9488 & $-$36.8146 & 4.1 & 4.1 & 0.12 & 2.96 & WISEA J045900.06$-$324223.8 \\
\hline Pulsar \\ \hline
ASKAP~J092430.8$-$581407 & 141.1283 & $-$58.2352 & 278.3944 & $-$5.5956 & 27.8 & 12.6 & 0.16 & 6.48 & PSR B0923$-$58 \\
ASKAP~J102438.7$-$071919 & 156.1611 & $-$7.3218 & 251.7013 & $+$40.5156 & 74.9 & 21.3 & 0.45 & 3.37 & PSR J1024$-$0719 \\
ASKAP~J164622.6$-$440543 & 251.5941 & $-$44.0953 & 340.8074 & $+$0.8271 & 126.3 & 4.8 & 0.75 & 4.16 & PSR J1646$-$4405 \\
ASKAP~J165817.0$-$430649 & 254.5708 & $-$43.1136 & 342.9345 & $-$0.2145 & 9.1 & 5.0 & 0.64 & 1.21 & PSR J1658$-$4306 \\
ASKAP~J170941.4$-$440113 & 257.4223 & $-$44.0203 & 343.4702 & $-$2.4061 & 19.8 & 12.6 & 1.95 & 0.51 & PSR J1709$-$4401 \\
ASKAP~J191049.2$-$532057 & 287.7049 & $-$53.3492 & 343.6519 & $-$24.2492 & 10.4 & 5.2 & 0.67 & 0.93 & PSR J1910$-$5320 \\
ASKAP~J210154.9$-$480158 & 315.4787 & $-$48.0327 & 351.4858 & $-$41.2883 & 4.5 & 4.9 & 0.15 & 2.54 & PSR J2101$-$4802 \\
ASKAP~J212945.1$-$042910 & 322.4377 & $-$4.4860 & 48.9104 & $-$36.9392 & 7.3 & 4.6 & 0.60 & 0.78 & PSR J2129$-$0429 \\
ASKAP~J214435.6$-$523708 & 326.1483 & $-$52.6189 & 343.4344 & $-$47.1233 & 9.1 & 9.1 & 0.40 & 1.33 & PSR J2144$-$5237 \\
\hline Star \\ \hline
ASKAP~J040003.7$-$290216 & 60.0155 & $-$29.0378 & 227.0214 & $-$48.6294 & 4.2 & 6.0 & 0.55 & 0.69 & HD 25284 \\
ASKAP~J040711.4$-$291834 & 61.7974 & $-$29.3094 & 227.7951 & $-$47.1377 & 6.8 & 12.2 & 1.18 & 0.36 & 1RXS J040710.6$-$291823 \\
ASKAP~J051629.6$-$313543 & 79.1233 & $-$31.5953 & 234.6099 & $-$32.9590 & 16.7 & 13.4 & 1.15 & 0.56 & UCAC4 293-006715 \\
ASKAP~J051921.2$-$072049 & 79.8384 & $-$7.3470 & 208.9808 & $-$23.7972 & 6.8 & 8.5 & 0.54 & 0.80 & HD 34736 \\
ASKAP~J141215.7$+$022433 & 213.0654 & $+$2.4092 & 344.4263 & $+$58.6104 & 30.4 & 23.2 & 0.62 & 1.99 & CU Vir \\
ASKAP~J191702.2$-$523856 & 289.2590 & $-$52.6488 & 344.6343 & $-$25.0026 & 9.8 & 10.1 & 0.89 & 0.65 & L274-89 \\
\hline Unidentified \\ \hline
ASKAP~J045522.4$-$350523 & 73.8435 & $-$35.0897 & 237.7324 & $-$37.9780 & 41.4 & 17.9 & 0.88 & 1.57 & - \\
ASKAP~J212758.7$-$470528 & 321.9947 & $-$47.0910 & 352.1271 & $-$45.7595 & 8.3 & 8.7 & 1.02 & 0.50 & - \\
\hline LPT \\ \hline
ASKAP~J165130.3$-$450520 & 252.8761 & $-$45.0888 & 340.6336 & $-$0.5109 & 12.7 & 15.6 & - & - & - \\
ASKAP~J170036.6$-$445758 & 255.1525 & $-$44.9661 & 341.7369 & $-$1.6884 & 8.0 & 10.0 & 1.87 & 0.32 & - \\
\bottomrule
\end{tabular}

\label{tab:transients}
\end{table*}

\begin{figure*}[t!]
\centering
\includegraphics[width=\linewidth]{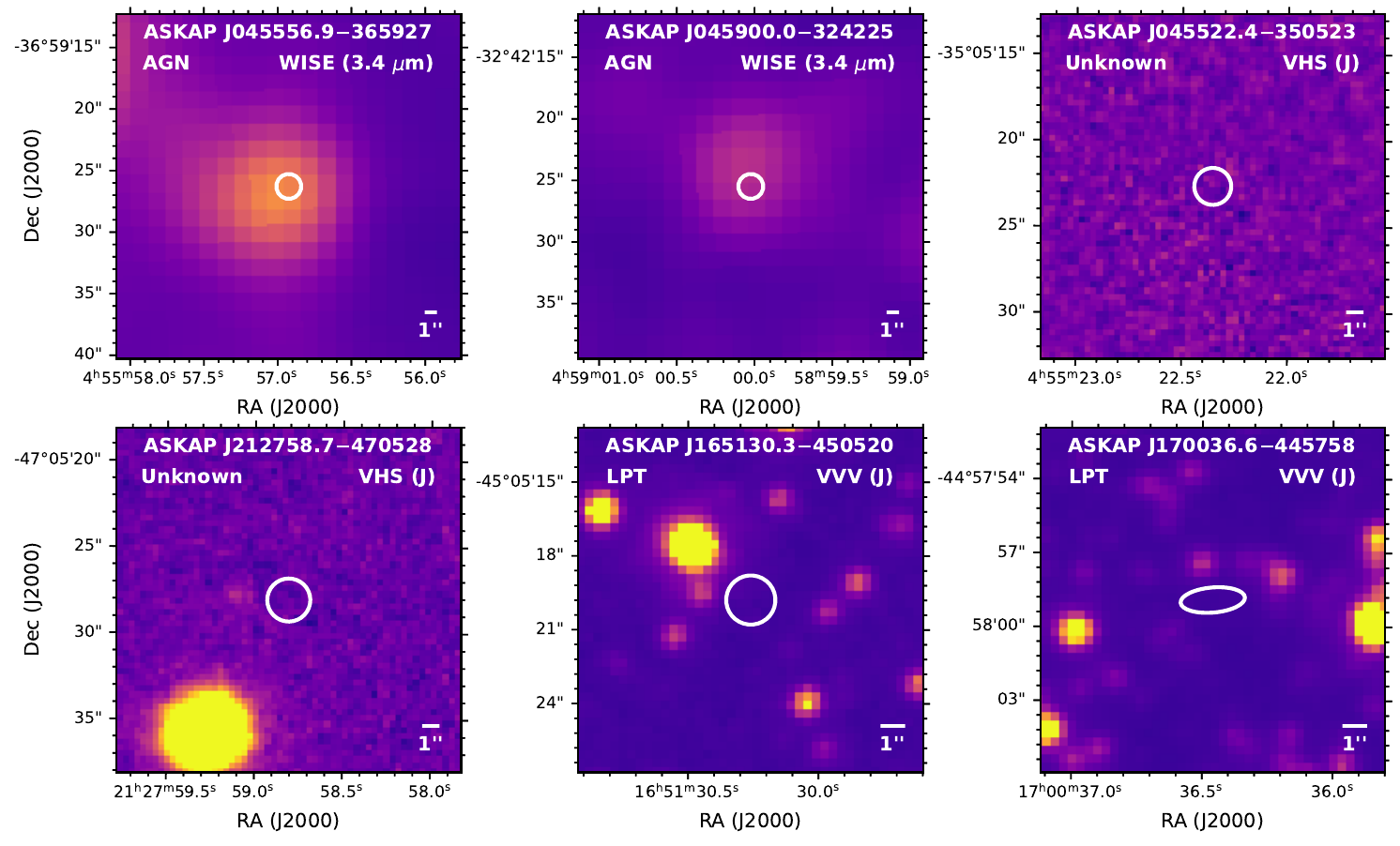}
\caption{Optical images overlaid with the $1\sigma$ positional uncertainty regions of the radio sources requiring further multi-wavelength searches. For each panel, the source name, classification, optical/infrared survey, and observing band are shown at the top. The angular scale is shown in the lower right corner. The $1\sigma$ localisation region of the radio source is given by the white ellipse. The positional uncertainties are derived from ASKAP observations (with added 1\,arcsec uncertainty floor to account for potential \updated{systematic astrometric errors}), except for \lpttwo{}, which is derived from the MeerKAT observation. See discussions in Section~\ref{sec:results}. }
\label{fig:optical_overlay}
\end{figure*}

The results presented in this paper come from the first two weeks of operation, from \updated{11~July~2025 to 26~July~2025}. 
During this period the VASTER system ran on 25 EMU observations, and two Target of Opportunity \updated{(ToO)} observations, while we tested the system stability. 
The total observing time was 205\,h.
These observations were conducted at a central frequency of 943.5\,MHz with a bandwidth of 288\,MHz. The CRACO system was not operating during this period, so no millisecond-timescale searches are available.
Details of the observations are given in Table~\ref{tab:survey}.
The sky coverage of these observations is shown in Figure~\ref{fig:survey_map}.

The VASTER system detected a total of 399 candidates during the first two weeks of operation.
We reviewed each candidate manually using the inspection tool.
Most were found to be false detections caused by
(i) antenna bandpass variations and time-dependent antenna amplitude gain changes (as a result of antenna failures or system changes);  
(ii) direction-dependent phase errors (as a result of extreme ionospheric conditions); and 
(iii) antenna pointing errors. 
\updatednew{These effects can produce non-PSF-like residual artefacts around field sources after sky-model subtraction, which can subsequently be misinterpreted as transient events}. 
These artefacts are particularly strong around bright sources and, in the case of pointing errors, \updated{result} in increased errors towards the beam edge. 
As these residual PSF artefacts rotate with time and/or increase in amplitude, they can trigger false detections at the source location or even some distance from the source if the residual PSF is particularly extended \updated{(see \ref{sec:vaster_outputs} for example artefacts)}.

After manual inspection of the candidates, we found 21 real transients. 
Their radio light curves are shown in Figure~\ref{fig:lightcurves}. 
We identified multi-wavelength \updated{counterparts} for most of these sources using our inspection tool which incorporates information from external databases; these include known pulsars and stars. 
For transients without an obvious optical/IR counterpart, we \updated{performed additional manual checks using} the VizieR catalogue\footnote{\url{https://vizier.cds.unistra.fr/viz-bin/VizieR}} and the European Southern Observatory archive\footnote{\url{https://archive.eso.org/scienceportal/home}} to find any fainter counterparts \updated{not included in SIMBAD}. 
ASKAP typically achieves sub-arcsec localisation accuracy for mJy-level sources. 
However, since it is not phase referenced, ASKAP can have \updated{systematic astrometric} errors of about 0.6\,arcsec in RA and 0.4\,arcsec in Dec \citep{McConnell2020PASA...37...48M}. 
To account for this, we used a conservative approach by adding 1\,arcsec uncertainty floor to the statistical uncertainties when cross-matching with optical surveys. 
Figure~\ref{fig:optical_overlay} shows their $1\sigma$ positional uncertainty regions overlaid on optical images.   
We found four sources with no multi-wavelength counterpart. 

Based on these investigations, the transients were classified as: nine known pulsars; six radio stars; two scintillating AGN; two new long period transients and two as-yet unidentified sources. 
Their properties are summarised in Table~\ref{tab:transients}, and their sky positions are plotted in Figure~\ref{fig:survey_map}. 
We briefly discuss individual sources in the following subsections.

\subsection{Long period transients}
LPTs are known to be highly intermittent in their activity \citep[e.g.,][]{Hurley-Walker2022Natur.601..526H}. 
One of the goals of VASTER is to detect LPTs in \updated{real time},
to enable rapid follow-up before they become inactive.

We detected two new LPTs in the first two weeks of VASTER real-time operations. These were both in the Galactic plane field, SB74690. 
We detected two pulses from each source in the initial observation. 
On this basis we conducted two ASKAP follow-up observations $\sim$10 days later (SB74883 and SB74955) and detected more pulses, confirming their periodicity. 
We also conducted two 3\,h MeerKAT observations in the UHF band (544--1087\,MHz) three months after discovery and detected one source at the expected time.\footnote{The MeerKAT follow up was slightly delayed because we focused on system testing \updated{in} the initial phase of VASTER operation. In future we aim for more rapid follow ups for unusual objects.}
The results and interpretation for each object are discussed in the following subsections. 


\subsubsection{LPT: \lptone}

\begin{figure}[t!]
\centering
\includegraphics[width=\columnwidth]{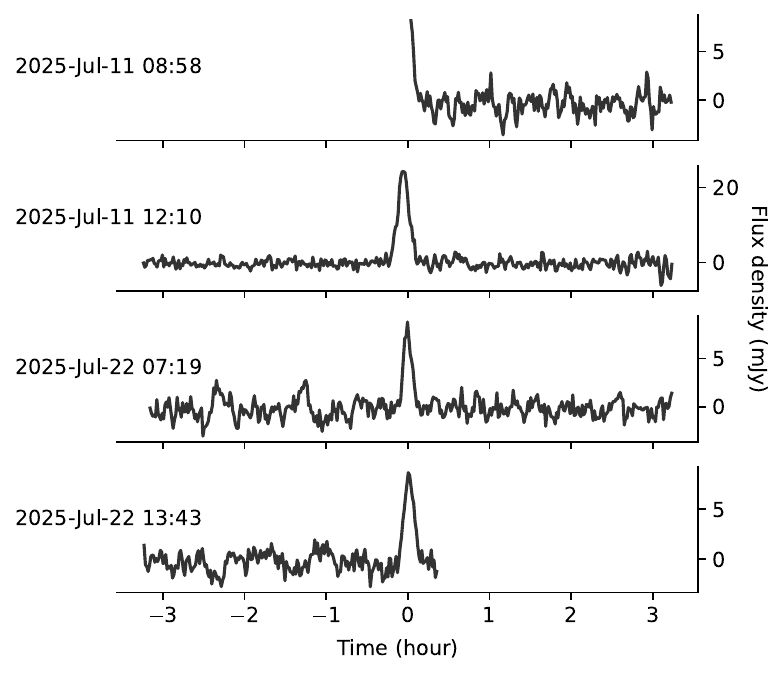}
\caption{Four detected pulses of \lptone{} aligned to its measured period. The observation start times are listed on the left of each detection. The light-curve time resolution is averaged to 1\,min. }
\label{fig:lc_lpt1}
\end{figure}

\begin{figure}[t!]
\centering
\includegraphics[width=\columnwidth]{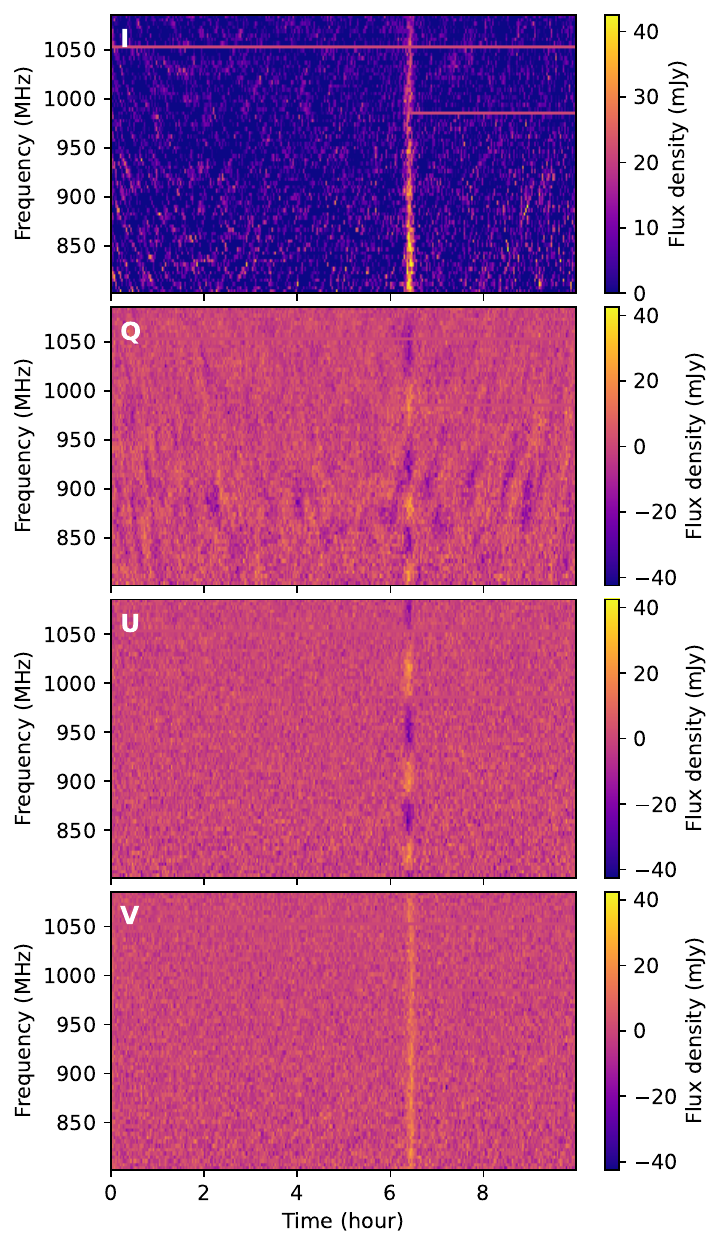}
\caption{Full-Stokes dynamic spectrum of \lptone{} in observation on \updated{11~July~2025}. The time resolution is averaged to 1\,min, and the frequency resolution is averaged to 4\,MHz. This observation used an experimental field-based calibration scheme to reduce calibration overheads involved when switching between bands and/or footprints, and therefore resulted in an excess of residual leakage from Stokes I to Stokes Q, other Stokes parameters were not adversely affected. }
\label{fig:ds_lpt1}
\end{figure}

\updated{
We detected four pulses across three ASKAP observations.
To characterise their properties, we extracted full-polarisation dynamic spectra at the highest available time and frequency resolution (10\,s and 1\,MHz) from the ASKAP visibilities with DStools \citep{Pritchard2025zndo..15232974P}. 
At this resolution, we do not observe any short-term substructure that could be used to measure a dispersion measure.
Figure~\ref{fig:lc_lpt1} shows the light curves obtained by averaging the total intensity dynamic spectra over frequency and binning the time resolution from 10\,s to 1\,min.
Figure~\ref{fig:ds_lpt1} shows the full-polarisation dynamic spectra for the brightest pulse, averaged to a resolution of 1\,min and 4\,MHz.
}

The times of arrival for three of the pulses, along with their uncertainties, were measured by fitting multi-component Gaussian functions to their pulse profiles. 
We estimated a period of $6.4772\pm0.0002$\,h with tempo2 \citep{Hobbs2006MNRAS.369..655H}. 
In the timing model, the source position was fixed at the value derived from ASKAP imaging, and we fit only for the spin frequency. 

To measure spectral index, we integrated the on-pulse phase (where the S/N $>6$) and binned the data into 16\,MHz frequency channels across the total 288\,MHz bandwidth.  
We measured a steep spectral index of $\sim-2.8$ for the brightest pulse \updated{(on 11~July~2025)}, while the spectral index changed to be $\sim-1.0$ for the fainter pulses \updated{(on 22~July~2025)}. 
We estimated a rotation measure (RM) of $-141\pm3$\,rad\,m$^{-2}$ for the brightest pulse \updated{with RM-Tools \citep{VanEck2026ApJS..283...28V}}. 
After applying RM correction, the linear polarisation fraction is $\sim70\%$. 
This object is also highly circularly polarised $\sim45\%$ (see Figure~\ref{fig:ds_lpt1}). 
The fainter pulses on \updated{22~July~2025} show a consistent RM of $-139\pm6$\,rad\,m$^{-2}$. 
We measured the pulse width (at FWHM) \updated{as} $\sim570\pm15$\,s, corresponding to a duty cycle of $\sim2.4\%$. 
This is comparable to that of the previously discovered LPT ASKAP~J183950.5$-$075635.0 \citep{Lee2025NatAs...9..393L}, which has a similar rotation period.

The field containing \lptone{} has been observed in 61 archival ASKAP observations, including the VAST survey, spanning from April~2019 to November~2024.
Since the end of 2022, the VAST survey has monitored this field approximately every two weeks.
The source was detected in only one of these 61 observations, a RACS observation on \updated{21~November~2024}.
During the detection, the source had a flux density of $\sim$4\,mJy and exhibited strong circular polarisation at $\sim$60\% level.
Using the current timing solution, the RACS detection corresponds to a pulse phase of approximately $-0.1$, slightly offset from the predicted arrival time.
The RACS observation is only 15 minutes long, and the pulse is too faint to allow a measurement of the arrival time.
Additional detections from future follow-up observations would be useful to refine the timing solution and resolve this discrepancy.

We observed this object with MeerKAT on \updated{10~October~2025} at UHF band (544--1087\,MHz) for 3\,h but detected no emission. 
This is not surprising as this object \updated{exhibits} rapid flux decay: the pulse peak flux density \updated{decreased} from 20\,mJy to 5\,mJy within 10 days, and it is possible that this source \updated{had dropped} below the MeerKAT sensitivity limit ($3\sigma$ limit of $\sim$0.1\,mJy) after three months. 

We do not find any multi-wavelength counterpart for this object (see Figure~\ref{fig:optical_overlay}). 
The deepest infrared observation available in this region is the VISTA Variables in the Via Lactea (VVV) survey \citep{Minniti2010NewA...15..433M}, which provides a $5\sigma$ limiting magnitude of $J>20.5$\,mag (AB). 
Given the low Galactic latitude and high extinction along this line of sight ($\sim$10\,mag at $J$ band; \cite{Schlafly2011ApJ...737..103S}), this limit is less constraining than other LPTs for which multi-wavelength counterparts have been detected. 


\subsubsection{LPT: \lpttwo}

\begin{figure}[t!]
\centering
\includegraphics[width=\columnwidth]{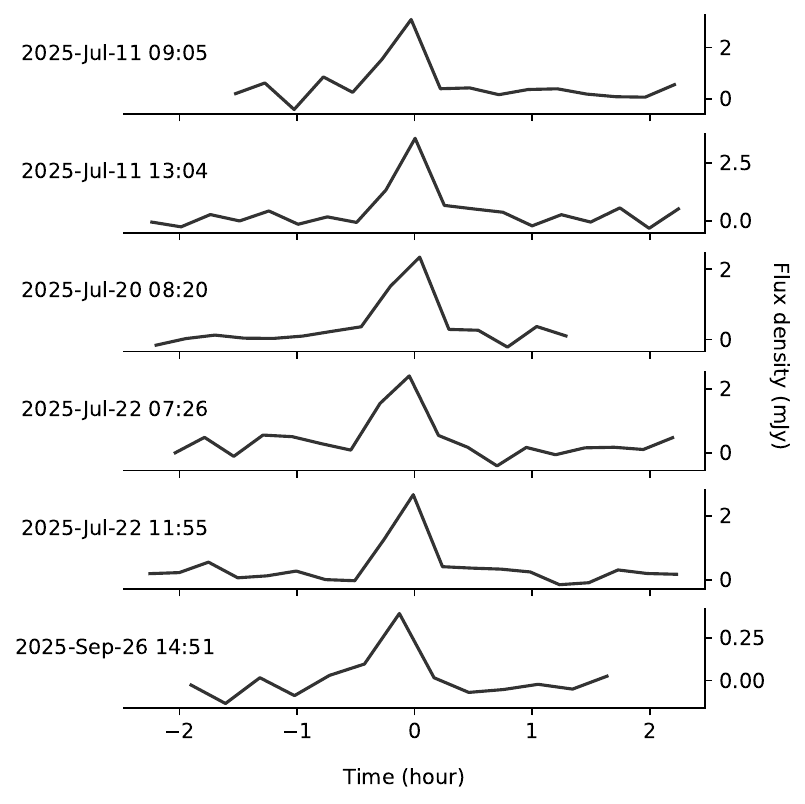}
\caption{Six detected pulses of \lpttwo{} aligned to its measured period. The first five pulses were detected with ASKAP, and the last pulse was detected with MeerKAT. The observation start times are listed on the left of each detection. The light-curve time resolution is averaged to 15\,min.  }
\label{fig:lc_lpt2}
\end{figure}

\updated{We detected five pulses across three ASKAP observations. 
We extracted full-polarisation, high-resolution dynamic spectra using the same method as above. 
Figure~\ref{fig:lc_lpt2} shows the light curves obtained by averaging the total intensity dynamic spectra over frequency and binning the time resolution from 10\,s to 15\,min.}

This object was not detected in any of the 115 ASKAP archival observations from April~2019 to November~2024.
We measured \updated{a} pulse width of $20\pm3$\,minutes, corresponding to a duty cycle of approximately $7\%$. 
The dynamic spectra show weak circularly polarised emission ($\sim$30\% in Stokes~V), but no detectable linear polarisation in Stokes~Q or Stokes~U.

We conducted a 3\,h MeerKAT follow-up observation at UHF band on 26~September~2025, and detected a very faint pulse at the expected time (see Figure~\ref{fig:lc_lpt2}). 
The decreased flux density suggests rapid fading for this object (\updatednew{dropping} from $\sim$2.5\,mJy to $\sim$0.25\,mJy), similar to other LPTs \citep[e.g.,][]{Hurley-Walker2022Natur.601..526H}. 
The MeerKAT observation provides an independent position measurement of RA = $17^{h}00^{m}36.46^{s}\pm0.91$\,arcsec and Dec = $-44^\circ57'58.91''\pm0.59$\,arcsec, consistent with the ASKAP position. 

We estimated the period with all six detected pulses (including the fainter one in the MeerKAT observation) using the same method as above, and the best measured period is $4.6933\pm0.0002$\,h. 
We do not identify any multi-wavelength counterpart for this object (see Figure~\ref{fig:optical_overlay}).
The VVV survey provides a $5\sigma$ limiting magnitude of $J>19.8$\,mag \citep{Minniti2010NewA...15..433M}. 
The extinction along this line of sight is $\sim$2\,mag at $J$ band \citep{Schlafly2011ApJ...737..103S}.  

\subsection{Unidentified sources}

Two transients lack multi-wavelength counterparts and remain of uncertain nature.
We analysed their polarisation properties, checked radio archival observations, and conducted follow-up observations, to investigate their nature.

\begin{figure}[t!]
\centering
\includegraphics[width=\columnwidth]{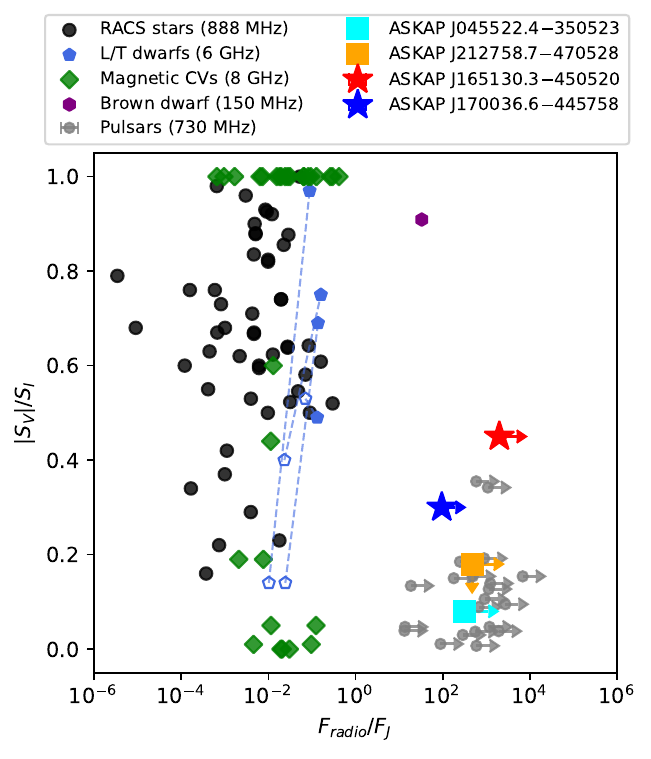}
\caption{Fractional circular polarization versus radio-to-near-infrared flux ratio for stellar objects and pulsars. 
The red and blue stars represent the two LPTs detected in this work, and the cyan and orange squares represent the two unidentified sources detected in this work. 
The black circles represent radio stars measured in RACS at 888\,MHz~\citep{Pritchard2021MNRAS.502.5438P}, the blue pentagons represent quiescence (open symbols) and peak (filled symbols) auroral emissions from L/T dwarfs at 6\,GHz~\citep{Kao2016ApJ...818...24K}, the green diamonds represent magnetic cataclysmic variables at 8\,GHz~\citep{Barrett2020AdSpR..66.1226B}, the purple hexagon represents the recently discovered brown dwarf BDR~J1750$+$3809 at 150\,MHz~\citep{Vedantham2020ApJ...903L..33V}, the gray circles represent a group of pulsars measured at 730\,MHz~\citep{Dai2015MNRAS.449.3223D}. 
We note that the two unidentified sources are consistent with the pulsar region. 
Figure adapted from \cite{Wang2022ApJ...930...38W}. 
}
\label{fig:stokes_v_infrared}
\end{figure}

\subsubsection{Pulsar candidate: \unkone}

\unkone{} was detected in 11 out of 29 archival ASKAP observations, with flux densities ranging from 1.5 to 2\,mJy in the detected epochs.
This object is circularly polarised with a Stokes~V flux density of $-135\pm24\,\mu$Jy, corresponding to \updated{circular} polarisation fraction of $8\%$. 
Circular polarisation is rare in radio sources, and \updated{is typically} associated with pulsars or stellar flares, which often show fractional circular polarisation of \updatednew{$\sim$10\% or above} \citep[e.g.,][]{Lenc2018MNRAS.478.2835L,Pritchard2021MNRAS.502.5438P}. 
We do not find any multi-wavelength counterparts for this object (see Figure~\ref{fig:optical_overlay}), leading to a $5\sigma$ limiting magnitude $J>21.3$\,mag from the Vista Hemisphere Survey (VHS; \cite{McMahon2013Msngr.154...35M}). 
The combination of high fractional circular polarisation and the absence of an infrared counterpart suggests that \unkone{} is likely a pulsar candidate (as shown in Figure~\ref{fig:stokes_v_infrared}). 

We observed \unkone{} with the Australia Telescope Compact Array (ATCA) on \updated{4~November~2025} for 4.5\,h \updated{between} 1140--3060\,MHz (project code C3363). 
The primary aim was to measure the spectral index. 
We used PKS J1934$-$638 as the primary bandpass calibrator, and PKS 0439$-$337 as the phase calibrator. 
The data were flagged, calibrated, and imaged with CASA \citep{CASATeam2022PASP..134k4501C}. 
However, no emission was detected in the image, resulting in a $3\sigma$ upper limit of $16\,\mu$Jy. 

We observed \unkone{} with the Parkes telescope on 30~November~2025 for 45\,min (project code PX143). 
The observation covered a frequency range of 704--4032\,MHz \citep{Hobbs2020PASA...37...12H}, divided into 3328 channels with a bandwidth of 1\,MHz each. 
Pulsation searches were performed using the PRESTO software \citep{Ransom2002AJ....124.1788R}. 
To maintain sensitivity to short–orbital-period pulsars in binary systems, we carried out a Fourier-domain acceleration search with a drift-rate range of $z=\pm150$ \citep{Andersen2018ApJ...863L..13A, Ng2015MNRAS.450.2922N}. 
The search was conducted over a dispersion measure (DM) range of 0--100\,pc\,cm$^{-3}$ (the estimated maximum DM at this direction is $\sim$40\,pc\,cm$^{-3}$; \cite{Cordes2002astro.ph..7156C,Yao2017ApJ...835...29Y}) using the RFI-cleaned 960--3008 MHz sub-band \citep{Zhang2019ApJ...885L..37Z}. No pulsed emission was detected. Assuming a detection threshold of ${\rm S/N}=8$, the radiometer equation \citep{Lorimer2004hpa..book.....L} yields a flux-density upper limit of $\sim$27~$\mu$Jy at a central frequency of $f_c=1984$ MHz, for an intrinsic duty cycle of 10\% and spin periods \updatednew{of $\sim$10\,ms or longer}.

The non-detections with both ATCA and Parkes are likely attributed to the highly intermittent nature of this object.



\subsubsection{Pulsar candidate: \unkthree}
\updated{\unkthree{} is detected in 19 of 89 archival ASKAP observations. The Stokes I integrated flux density varies between 0.98 to 6.4 mJy.}
We do not detect any circular polarisation for this object with a $3\sigma$ limit of $\sim$0.1\,mJy, corresponding to a polarisation fraction $<18\%$. 
\updated{This object was first detected as a transient with unknown progenitor in the VAST pilot survey \citep{Murphy2021PASA...38...54M}. Dynamic spectra of the brightest archival detections show similar rising and falling behaviour as seen in Figure \ref{fig:lightcurves} on timescales of minutes.}

We do not find any multi-wavelength counterparts for this object (see Figure~\ref{fig:optical_overlay}), leading to a $5\sigma$ limiting magnitude $J>21.1$\,mag from VHS \citep{McMahon2013Msngr.154...35M}. 
\updatednew{We identified a radio counterpart, \unkthreegx{}, in the low-frequency GLEAM-X survey,} where it exhibits flux densities of 38\,mJy at 76\,MHz and 4\,mJy at 227\,MHz \citep{Ross2024PASA...41...54R}.
The inferred steep spectral index ($\sim-1.4$), together with the strong variability seen in the ASKAP observations, suggests that \unkthree{} is likely a pulsar candidate.
This object has been detected by the CRACO system, and the details will be presented in Jaini et al. (in prep). 

\subsection{Other transient and variable sources}
The remaining transients all have identified counterparts, as listed in Table~\ref{tab:transients}.

\subsubsection{AGN}

Two transients are classified AGN based on their infrared counterparts in the WISE survey (see Figure~\ref{fig:optical_overlay}) and their locations on the WISE colour–colour diagram \citep{Wright2010AJ....140.1868W}.
At this observing frequency ($\sim$1\,GHz) and timescale (15\,min), their variability is unlikely to be intrinsic, as the estimated brightness temperature would greatly \updated{exceed} the inverse Compton limit $T_b\sim10^{12}$\,K for incoherent synchrotron emission \citep{Kellermann1969ApJ...155L..71K}. 
Instead, the most plausible origin of the observed variability is enhanced scintillation caused by nearby plasma screens \citep{Dennett-Thorpe2002Natur.415...57D}. 
AGN exhibiting this behaviour are commonly referred to as intra-day variables (IDVs). 
We note that the two IDVs are found in the same ASKAP field, with a separation of only a few degrees. 
This clustered sky distribution is consistent with that reported by \cite{Wang2023MNRAS.523.5661W}.

\subsubsection{Known pulsars}

We identified nine pulsars by cross-matching with the ATNF Pulsar Catalogue \citep{Manchester2005AJ....129.1993M}.
\updated{Pulsars are known to exhibit variability due to a range of intrinsic and propagation effects.}
On this timescale (15\,min), the primary reason would be diffractive scintillation \citep[e.g.,][]{Wang2023MNRAS.523.5661W}, a propagation effect caused by small-scale fluctuations in the interstellar medium \citep{Narayan1992RSPTA.341..151N}. 
We estimated the diffractive scintillation timescales for these pulsars using their DMs and inferred distances \citep{Yao2017ApJ...835...29Y}.
Assuming a transverse velocity of 100\,km\,s$^{-1}$, we find that six of the nine pulsars have estimated diffractive timescales between 2 and 30 minutes at 1\,GHz, consistent with what we observed.
\updated{Pulsars} are also known to suddenly cease their radio emission for seconds, days, or years \citep{Kramer2006Sci...312..549K}. 
This intermittent behaviour can cause integrated flux density variations that can be detected by VASTER. 
PSR~J1709$-$4401 is likely a highly intermittent object, with \updated{clear `on' and `off' states} showing in its lightcurve (see Figure~\ref{fig:lightcurves}). 
Another object, PSR~J1646$-$4405 stands out as it is a candidate redback pulsar binary \citep{Zic2024MNRAS.528.5730Z}, and the variability is primarily due to eclipses by the companion. 

\subsubsection{Stars}

Three identified stars, HD~25284, 1RXS~J040710.6$-$291823, and L274-89 are K or M dwarf stars, which are
known to produce radio flares on timescales ranging from seconds to hours due to strong magnetic fields \citep[e.g.,][]{Dulk1985ARA&A..23..169D}. 
Among them, L274-89 has been reported in the Sydney Radio Star Catalogue \citep[][]{Driessen2024PASA...41...84D}, \updatednew{while 1RXS~J040710.6$-$291823 and HD~25284} have not been detected \updated{in the radio} before. 
Two of our identified stars, HD~34736 and CU Vir, are magnetic chemically peculiar stars. Both of them have been detected in radio before \citep{Semenko2024MNRAS.535.2812S,Driessen2024PASA...41...84D}. 
The remaining object, UCAC4 293-006715, has been classified \updatednew{as an} RS Canum Venaticorum (RS CVn) binary system \citep{GaiaCollaboration2023A&A...674A...1G}. 
RS CVn systems are known to have strong magnetic fields generated by rapid, tidally induced rotation periods, and are able to produce both quiescent and coherent radio bursts \citep[e.g.,][]{White1995ApJ...444..342W}. 
\updated{UCAC4 293-006715 has not previously been reported to exhibit radio emission.}

\section{Discussion}
\label{sec:discussion}


During the first two weeks of operation of the VASTER real-time system, we detected 21 highly variable sources over a total observing time of 205\,h,  corresponding to a rough detection rate of $\sim$1 transient per 10\,h. 
This rate is consistent with the previous work using the VASTER offline pipeline \updated{on a 15\,min timescale with the ASKAP pilot surveys} \citep{Wang2023MNRAS.523.5661W}. 

Among these 21 variable sources, four are unusual: two previously undetected LPTs and two pulsar candidates. 
A simple scaling suggests that VASTER is able to find $\sim$1 unusual transient per 50\,h of observing time. 
However, quantifying this discovery rate is difficult because the spatial distribution of LPTs (and pulsars) \updated{is} not isotropic. 
Instead, \citet{Dobie2024MNRAS.535..909D} note a strong bias towards them being located at very low Galactic latitudes. 
As shown in Figure~\ref{fig:survey_map}, excluding the ToO observations triggered by the analysis in this work, only two of the 25 observations analysed in this work (SB74690 and SB74878) are at low Galactic latitude. 
The only observation close to the Galactic centre (SB74690) contains both of the new LPTs, reinforcing the previously reported Galactic latitude dependence. 

For the two discovered LPTs, the \updated{real-time} detection enabled by VASTER allowed us to conduct follow-up observations rapidly within $\sim$10 days, successfully re-detecting both objects.
A further follow-up observation three months later showed that one object, \lptone{}, had become inactive, and the other one, \lpttwo{}, was $\sim$10 times fainter than before. 
\updated{This highlights the importance of VASTER's ability to run in real time, enabling timely} follow-up observations for highly intermittent sources such as LPTs. 

VASTER is currently running on 15\,min timescales, making it most sensitive to variability on timescales of hours (such as LPTs with periods of a few hours like those reported in this work). 
\updatednew{In future, we plan to move to higher time resolution, and ultimately perform search at the raw correlator dump rate (10\,s).}
This would allow us to detect objects on \updated{seconds-to-minutes timescales}, including LPTs with periods of a few minutes. 
We also plan to improve candidate classification strategies to reduce the number of false detections. 
In this work, a total of 399 candidates were reported by VASTER, of which 58 \updated{were found to be} genuine detections after manual inspection (corresponding to 21 unique transients since many were detected multiple times in adjacent beams and/or different fields), implying a true-positive rate of $\sim$15\%. 
While the current number of candidates is manageable, \updated{classification still requires some human effort (and time)}. 
As the volume of data (and human-labelled candidates) grows, VASTER will provide an ideal training set for machine-learning–based classification, which will significantly reduce human workload and improve system robustness.


This work represents the first demonstration of an image-domain transient detection pipeline operating commensally in real time on a widefield radio telescope, representing an important step towards transient searches with the SKA observatory \footnote{\url{https://www.skao.int/en}}. 
The bulk of LPTs reported to-date have been found via coarse fast imaging (similar to that reported in this work), circular polarisation searches, or as transients in single images and characterising them (periods, DM, RM, and pulse morphology) has required reprocessing the corresponding visibilities. 
This approach is not possible in the SKA era, as visibilities will not be stored due to data volume constraints. 
Hence, a robust real-time detection pipeline that can also automatically produce high-resolution spectra is necessary to continue \updated{scientific studies of LPTs and other transients in the SKA era}.

\section{Conclusion}

VASTER is the first short-timescale imaging and transient-detection pipeline operating commensally in real time on a wide-field radio telescope.
In its first two weeks of operation, VASTER has demonstrated its capability to identify minute-cadence transients in real time, including the discovery of two new LPTs.
VASTER bridges the gap between existing transient surveys, which typically target either millisecond or day-to-year variability.
This capability makes ASKAP the only facility in the world with dedicated pipelines capable of discovering transients across a broad range of timescales: from milliseconds with CRAFT/CRACO, to seconds--hours with VASTER, and to days--years with the VAST survey.
With \updated{the} continued operation of ASKAP over the coming years, VASTER is expected to identify more LPTs and potentially new types of transients, providing a better understanding of the dynamic radio sky and \updatednew{its} diverse physical processes.

\section*{Data Availability}
The ASKAP data used in this paper can be accessed through the CSIRO ASKAP Science Data Archive
(CASDA\footnote{\url{https://data.csiro.au/domain/casdaObservation}}) under project codes AS113 and AS201. 
Among these observations, SB74815 and SB74873 were rejected by the EMU team, and SB74690 has not yet finished standard processing. 
These three observations are not publicly released. 

\section*{Acknowledgements}
 
We thank the anonymous referee for the careful review and constructive suggestions. 
YW acknowledges support through the Australian Research Council grant DP220102305 and FT190100155.
This research was supported by the Australian Research Council Centre of Excellence for Gravitational Wave Discovery (OzGrav), project number CE230100016. 
DLK was supported by NSF grants AST-1816492 and AST-2511757. 
RMS acknowledges support through ARC Discovery Project DP220102305.

This scientific work uses data obtained from Inyarrimanha Ilgari Bundara / the Murchison Radio-astronomy Observatory. We acknowledge the Wajarri Yamaji People as the Traditional Owners and native title holders of the Observatory site. CSIRO's ASKAP radio telescope is part of the Australia Telescope National Facility (\url{https://ror.org/05qajvd42}). Operation of ASKAP is funded by the Australian Government with support from the National Collaborative Research Infrastructure Strategy. ASKAP uses the resources of the Pawsey Supercomputing Research Centre. Establishment of ASKAP, Inyarrimanha Ilgari Bundara, the CSIRO Murchison Radio-astronomy Observatory and the Pawsey Supercomputing Research Centre are initiatives of the Australian Government, with support from the Government of Western Australia and the Science and Industry Endowment Fund.
The Australia Telescope Compact Array is part of the Australia Telescope National Facility (\url{https://ror.org/05qajvd42}) which is funded by the Australian Government for operation as a National Facility managed by CSIRO.
We acknowledge the Gomeroi people as the Traditional Owners of the Observatory site.
Murriyang, CSIRO's Parkes radio telescope, is part of the Australia Telescope National Facility (\url{https://ror.org/05qajvd42}) which is funded by the Australian Government for operation as a National Facility managed by CSIRO.
We acknowledge the Wiradjuri people as the Traditional Owners of the Observatory site.
The MeerKAT telescope is operated by the South African Radio Astronomy Observatory, which is a facility of the National Research Foundation, an agency of the Department of Science and Innovation.

Part of this work was performed on the OzSTAR national facility at Swinburne University of Technology. The OzSTAR program receives funding in part from the Astronomy National Collaborative Research Infrastructure Strategy (NCRIS) allocation provided by the Australian Government, and from the Victorian Higher Education State Investment Fund (VHESIF) provided by the Victorian Government. 
This work was supported by software support resources awarded under the Astronomy Data and Computing Services (ADACS) Merit Allocation Program. ADACS is funded from the Astronomy National Collaborative Research Infrastructure Strategy (NCRIS) allocation provided by the Australian Government and managed by Astronomy Australia Limited (AAL).
This research was supported by use of the Nectar Research Cloud, a collaborative Australian research platform supported by the NCRIS-funded Australian Research Data Commons (ARDC). 
This paper includes archived data obtained through the CSIRO ASKAP Science Data Archive, CASDA (\url{http://data.csiro.au}).
This research has made use of the SIMBAD database, operated at CDS, Strasbourg, France
This research has made use of the VizieR catalogue access tool, CDS, Strasbourg, France \citep{10.26093/cds/vizier}. The original description of the VizieR service was published in \citet{vizier2000}.
This paper draft has used ChatGPT for grammar checking. 

This research has made use of
Aegean~\citep{Hancock2012MNRAS.422.1812H,Hancock2018PASA...35...11H}, 
APLpy \citep{Robitaille2012ascl.soft08017R}, 
ASKAPsoft \citep{Guzman2019ascl.soft12003G}, 
Astropy~\citep{AstropyCollaboration2013A&A...558A..33A,AstropyCollaboration2018AJ....156..123A}, 
Astroquery \citep{Ginsburg2019AJ....157...98G}, 
Matplotlib \citep{Hunter2007CSE.....9...90H},
NumPy \citep{Harris2020Natur.585..357H}, 
oxkat\footnote{\url{https://github.com/IanHeywood/oxkat}}~\citep{Heywood2020ascl.soft09003H}, 
PyGDSM \citep{Price2016ascl.soft03013P}, 
SciPy \citep{Virtanen2020NatMe..17..261V}, 
Selavy \citep{Whiting2012PASA...29..371W}, 
tempo2 \citep{Hobbs2006MNRAS.369..655H}, 
and 
wsclean \citep{Offringa2014MNRAS.444..606O}.

\bibliography{bibtemplate}

\appendix

\begin{figure*}[t!]
\centering
\includegraphics[width=0.8\linewidth]{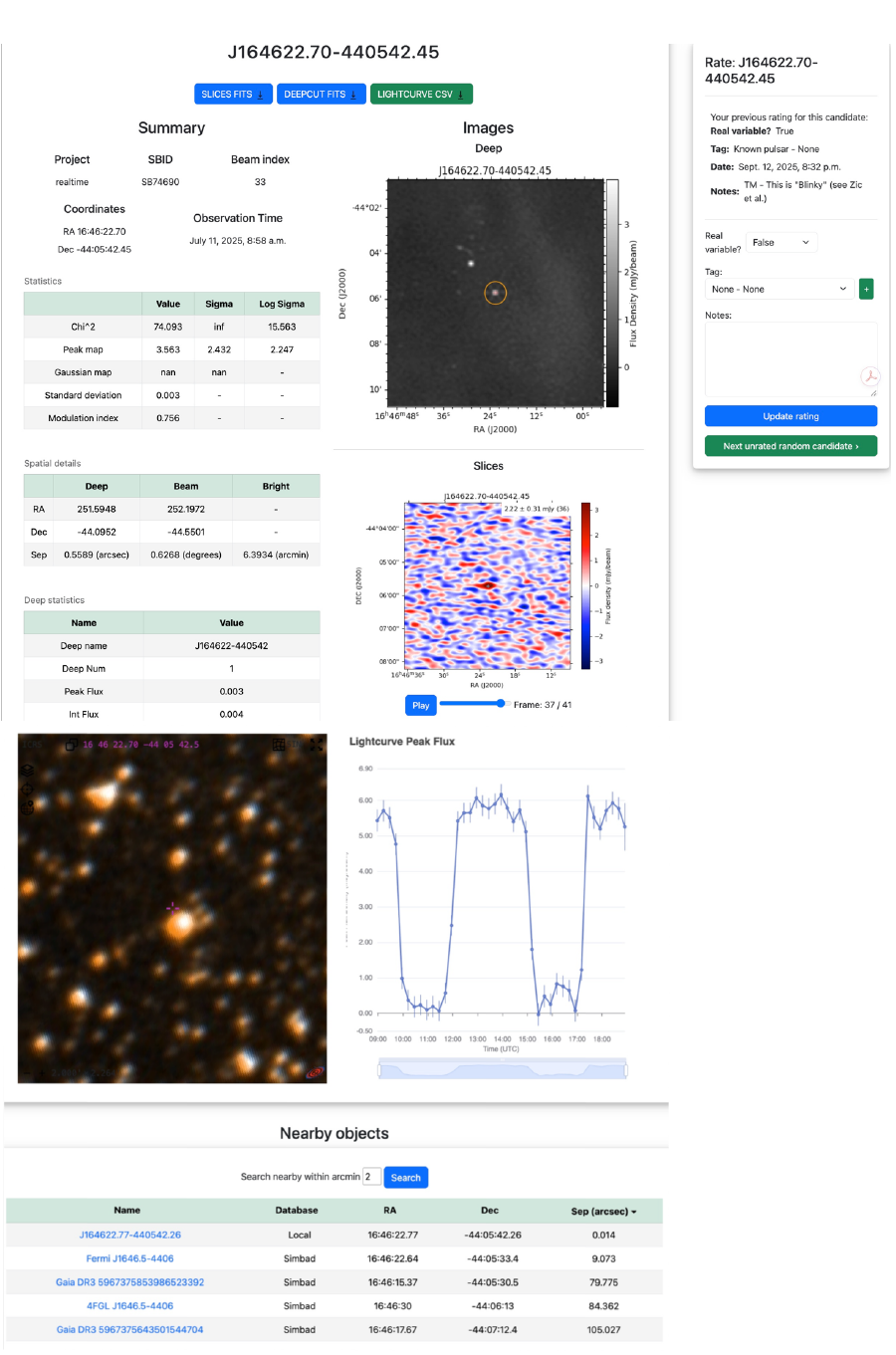}
\caption{
\updated{Example of a transient candidate displayed in the VASTER web inspection tool. 
The summary panel (top left) provides observation metadata and candidate statistics. 
The time-averaged deep image (top centre), the short-timescale residual images (`slices', middle), and the intra-observation light curve (bottom right) are shown as a central column. 
The right panel provides the classification interface, where users can assess if the candidate is real, assign a source type, and add notes. 
The bottom-left panel (next to the lightcurve) shows an Aladin interface, enabling visualisation of multi-wavelength images from major surveys. 
Additional information on nearby sources is provided in the lower panel. 
Note that the web interface is under active development, and future versions may differ from the one shown here. } 
}
\label{fig:webapp}
\end{figure*}

\begin{figure*}[t!]
\centering
\includegraphics[width=\linewidth]{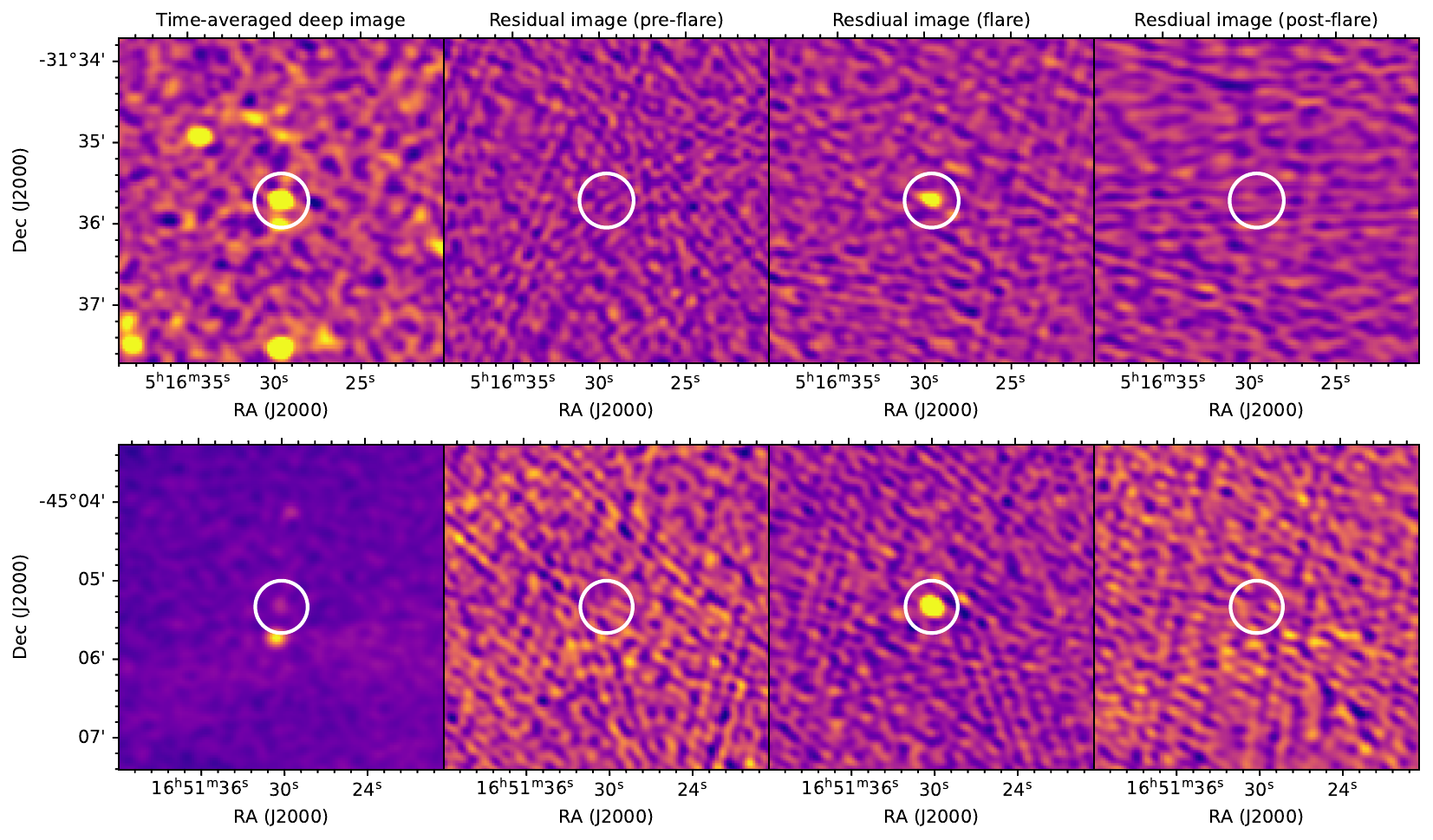}
\caption{
\updated{Examples of transient detections in the time-averaged deep images and short-timescale residual images, plotted from the VASTER outputs, for two sources: the flaring star ASKAP~J051629.6$-$313543 (top row) and the LPT ASKAP~J165130.3$-$450520 (bottom row).
From left to right in each row: the time-averaged deep image, the residual image prior to the flare, the residual image during the flare, and the residual image after the flare.
The white circle marks the target source position, with a radius of 20\,arcsec.
While the deep image contains multiple sources, these are effectively subtracted in the residual images; only the transient emission from the target source is visible during the flare.} 
}
\label{fig:example_flare}
\end{figure*}

\begin{figure*}[t!]
\centering
\includegraphics[width=\linewidth]{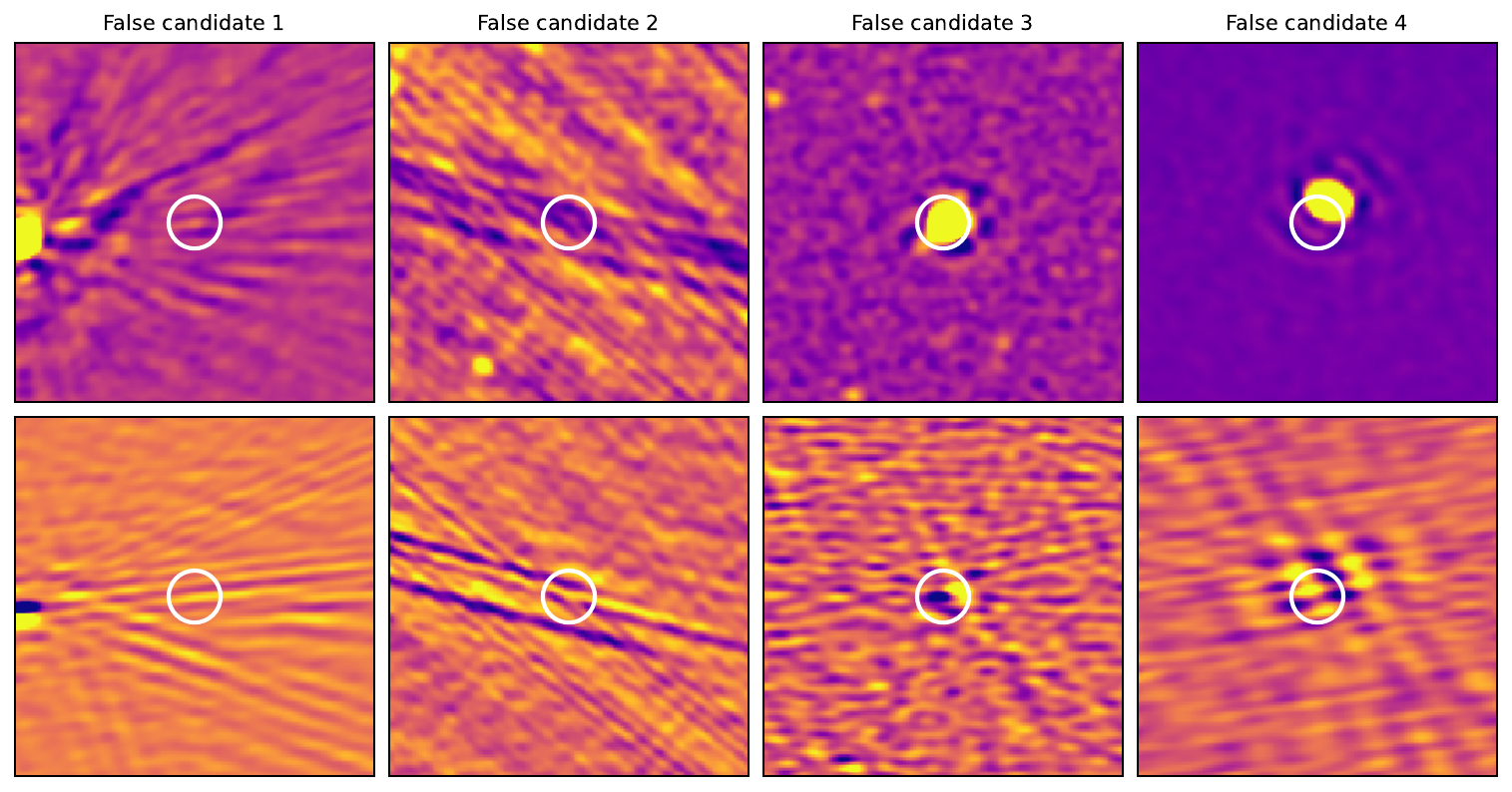}
\caption{
\updated{Examples of typical false candidates in the VASTER outputs.
Each column corresponds to one candidate: the top row shows the time-averaged deep images, and the bottom row shows the representative residual images.
The white circle marks the candidate source position, with a radius of 20\,arcsec. 
In the residual images, these candidates appear as non–PSF-like artefacts, which is a key indicator used to identify and reject false detections.} 
}
\label{fig:example_artefacts}
\end{figure*}

\updated{
\section{VASTER Candidate Examples}
\label{sec:vaster_outputs}

We present examples of VASTER outputs in this section. 
Figure~\ref{fig:webapp} shows an example candidate displayed in the VASTER web inspection tool. 
Figure~\ref{fig:example_flare} presents examples of real transient events in the time-averaged deep images and short-timescale residual images, showing how genuine sources appear in the VASTER outputs.
Figure~\ref{fig:example_artefacts} shows examples of typical false candidates, characterised by non–PSF-like artefacts in the residual images.
}

\end{document}